# Article Title:

# HEOM-QUICK2: a general-purpose simulator for fermionic many-body open quantum systems – An Update

## Authors:

**Daochi Zhang**

[Department of Chemistry, Fudan University, Shanghai 200433, China;

Hefei National Research Center for Physical Sciences at the Microscale, University of

Science and Technology of China, Hefei, Anhui 230026, China]

ORCID iD: orcid.org/ 0000-0003-1921-8141

**Lyuzhou Ye**

[Hefei National Research Center for Physical Sciences at the Microscale, University of

Science and Technology of China, Hefei, Anhui 230026, China]

ORCID iD: orcid.org/0000-0001-8016-722X

**Jiaan Cao**

[Hefei National Research Center for Physical Sciences at the Microscale, University of

| |
|---|
| Science and Technology of China, Hefei, Anhui 230026, China] |
| **Yao Wang** [Hefei National Research Center for Physical Sciences at the Microscale, University of Science and Technology of China, Hefei, Anhui 230026, China] |
| **Rui-Xue Xu** [Key Laboratory of Precision and Intelligent Chemistry, University of Science and Technology of China, Hefei 230026, China; Hefei National Research Center for Physical Sciences at the Microscale, University of Science and Technology of China, Hefei, Anhui 230026, China; Hefei National Laboratory, Hefei, Anhui 230088, China] |
| **Xiao Zheng\* (xzheng@fudan.edu.cn)** [Department of Chemistry, Fudan University, Shanghai 200433, China] ORCID iD: orcid.org/ 0000-0002-9804-1833 |
| **YiJing Yan** [Hefei National Research Center for Physical Sciences at the Microscale, University of Science and Technology of China, Hefei, Anhui 230026, China] |

## Conflict of Interest

There is no conflict of interest.

## Abstract

Many-body open quantum systems (OQSs) have a profound impact on various subdisciplines of physics, chemistry, and biology. Thus, the development of a computer program capable of accurately, efficiently, and versatilely simulating many-body OQSs is highly desirable. In recent years, we have focused on the advancement of numerical algorithms based on the fermionic hierarchical equations of motion (HEOM) theory. Being in-principle exact, this approach allows for the precise characterization of many-body correlations, non-Markovian memory, and non-equilibrium thermodynamic conditions. These efforts now lead to the establishment of a new computer program, HEOM for QUantum Impurity with a Correlated Kernel, version 2 (HEOM-QUICK2), which, to the best of our knowledge, is currently the only general-purpose simulator for fermionic many-body OQSs. Compared with version 1, the HEOM-QUICK2 program features more efficient solvers for stationary states, more accurate treatment of non-Markovian memory, and improved numerical stability for long-time dissipative dynamics. Integrated with quantum chemistry software, HEOM-QUICK2 has become a valuable theoretical tool for the precise simulation of realistic many-body OQSs, particularly the single atomic or molecular junctions. Furthermore, the unprecedented precision achieved by HEOM-QUICK2 enables accurate simulation of low-energy spin excitations and coherent spin relaxation. The unique usefulness of HEOM-QUICK2 is demonstrated through several examples of strongly correlated quantum impurity systems under non-equilibrium conditions. Thus, the new HEOM-QUICK2 program offers a powerful and comprehensive tool for studying many-body OQSs with exotic quantum phenomena and exploring applications in various disciplines.

**Graphical/Visual Abstract and Caption**

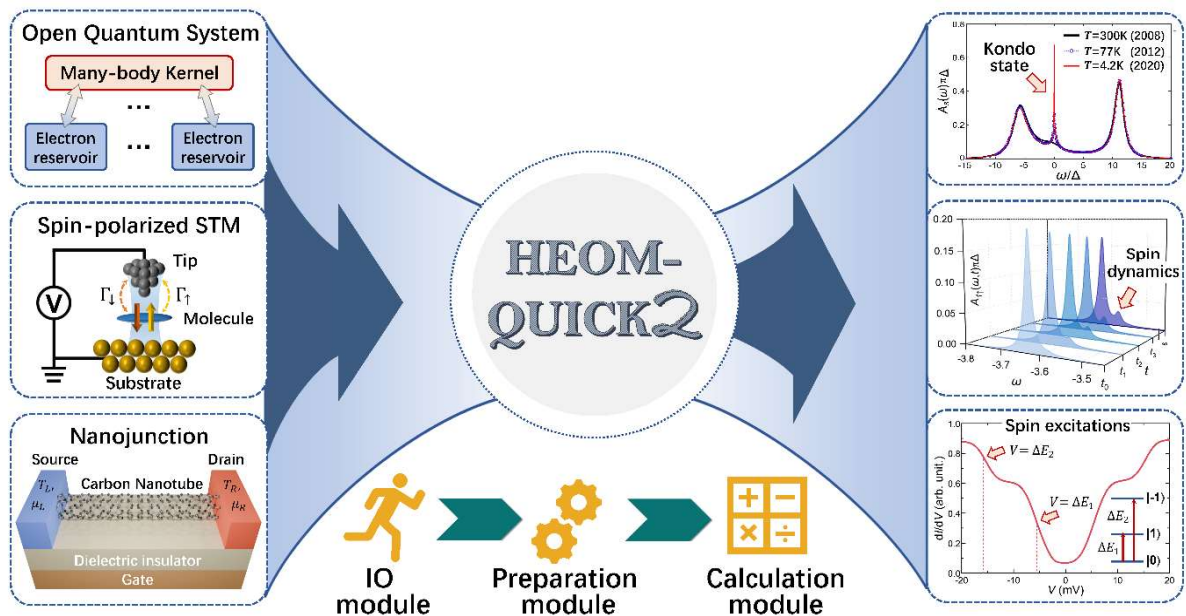

HEOM-QUICK2 is a general-purpose software package that enables accurately solving the stationary state and non-Markovian dynamics for fermionic many-body OQSs. Integrated with quantum chemistry software, HEOM-QUICK2 offers a valuable tool for simulating realistic many-body OQSs with an unprecedented precision.

# 1. INTRODUCTION

The concept of open quantum systems (OQSs) has found widespread applications in physics, chemistry, and biology. The intricate many-body interactions within OQSs, together with the dissipative coupling to the surrounding environment give rise to a diverse range of fascinating phenomena. Exploring the dynamics and stationary properties of many-body OQSs not only enhances our understanding but also provides valuable insights into the underlying physics mechanisms behind these phenomena.

Theoretical characterization of many-body OQSs relies on the accurate and efficient treatment of the many-body correlations ("correlation kernel") and the non-Markovian memory ("memory kernel"). Here, the non-Markovian memory means that the system states at present time depends on its history of evolution. The art of developing theoretical approaches for many-body OQSs has been refined over many years[1-8], encompassing the utilization of prominent methods such as the numerical renormalization group (NRG) method[9-14] and its time-dependent extension[15-19], the quantum Monte Carlo (QMC) method[20-31], the density matrix renormalization group (DMRG) method[32-36] and its time-dependent extension[37-43], the multiconfiguration time-dependent Hartree (MCTDH) method[44-47] and its multi-layer[48-52] and second-quantized version[53], the stochastic dynamical method[54-67], the real-time path integral (PI) method[68-75], and the steady-state density functional theory (i-DFT) method[76]. Despite the significant and remarkable success obtained by these advanced methods in the field of many-body OQSs, there are still some critical challenges to be addressed. As mentioned in Ref. [10], from the NRG point of view the multi-impurity quantum model presents a challenging task which requires additional numerical techniques[77] to achieve reliable and accurate dynamic simulations. The QMC method has to carefully tackle the fermionic sign problem which leads to a rapid growth in the required computational resources with simulation time, as discussed in Ref. [25, 30].

Over the past decade, the hierarchical equations of motion (HEOM) method first proposed by Tanimura and Kubo[78] has gained increasing attention in a broad interdisciplinary community[79-103] and has been extensively employed to treat many-body OQSs of practical interest. Particularly, the HEOM method has been employed to study real-time and steady-state quantum transport through nanojunctions[97,98,104-109], to examine the stability of molecular electronic devices[110-113], to account for various linear and nonlinear spectra in molecular and solid state materials[114-120], to evaluate charge and exciton transfer rates in biological systems[121-126], to simulate quantum ratchet processes and molecular motors in nanodevices[127,128], to unravel quantum information hidden in quantum entanglement states[129-132], and to investigate the essential problems in quantum thermodynamics[133-139], etc. Moreover, combined with quantum chemistry or first-principles approaches, the HEOM method allows for studies of correlated electronic structures in a wide range of realistic molecular or atomic systems[140-145].

The HEOM method enables accurately capturing the combined effects of non-Markovian memory and many-body correlations under nonequilibrium thermodynamic conditions and treats system dissipative dynamics in a non-perturbative manner. For generic many-body OQSs coupled to fermionic environments that satisfy Gaussian statistics (for example, noninteracting electron reservoirs), Yan et al. have proposed the formally exact fermionic HEOM in a compact form:[80]

$$\dot{\rho}^{(n)}_{j_1 \ldots j_n} = \left(-i\mathcal{L}_\text{s} + \sum_r \gamma_{j_r}\right)\rho^{(n)}_{j_1 \ldots j_n} + \sum_j \mathcal{A}_j\, \rho^{(n+1)}_{j j_1 \ldots j_n} + \sum_r \mathcal{C}_{j_r}\, \rho^{(n-1)}_{j_1 \ldots j_{r-1} j_{r+1} \ldots j_n}. \qquad (1)$$

Here, the auxiliary density operators (ADO) are denoted as $\rho^{(n)}_{j_1 \ldots j_n}$ with $n = 1, 2, \ldots$ and $j = 1, 2, \ldots$. It is noted that the anti-commutativity of fermions has been explicitly considered in the sequence of subscripts, for example, $\rho^{(3)}_{j_1 j_2 j_3} = -\rho^{(3)}_{j_2 j_1 j_3}$. Thus, the HEOM formalism effectively circumvents the fermionic sign problem by enumerating all possible sequences of ADOs. Specifically, $\rho^{(0)} = \rho_\text{S} \equiv \text{tr}_\text{B}[\rho_\text{T}]$ represents the reduced density operator (RDO) obtained by taking the partial trace of the system-environment total density matrix $\rho_\text{T}$ over the environment degrees of freedom. $n$ labels the tier of hierarchy and $\{j_n\}$ denote the principal dissipation modes (dissipaton)[146], respectively. The system Liouvillian is denoted as $\mathcal{L}_\text{S}$, and $\gamma_j$ represents the dissipation rate of the $j$th-principal dissipation mode. The dissipation superoperators $\mathcal{A}_j$ and $\mathcal{C}_{j_r}$ will be elaborated on in Sec. 2.2, where their detailed forms will be provided. Apparently, while the system Hamiltonian determines the Liouvillian $\mathcal{L}_\text{S}$ and the size of matrices representing the RDO/ADOs, the dissipation rates and superoperators depend on the statistical properties of environments. Thus, without the aid of extra techniques, the HEOM method is in principle capable of accurately describing the dissipative dynamics of many-body

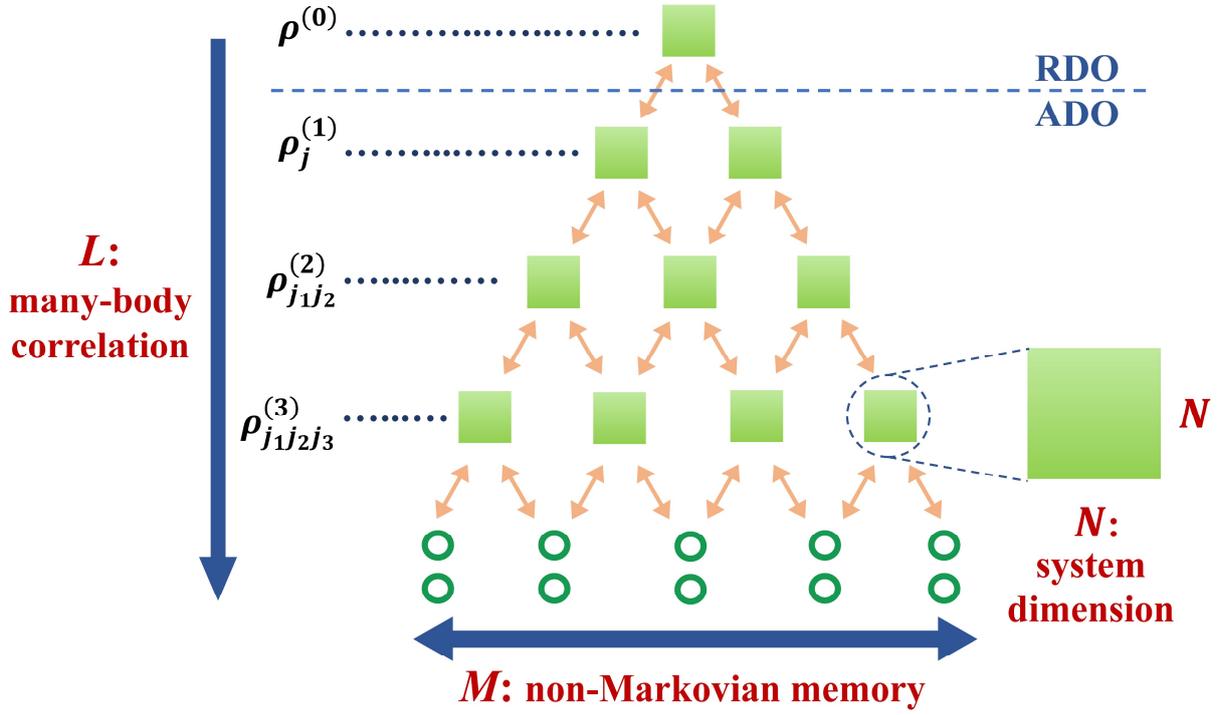

Figure 1: The schematic diagram for the hierarchical structure of RDO $\rho^{(0)} = \rho_S$ and ADO $\rho^{(n)}_{j_1\cdots j_n}$ (green squares). The structure of hierarchy extends into the horizontal (width) and the vertical (height) dimensions. In practice, the hierarchy has to be truncated at a finite size of the horizontal and vertical dimensions, denoted by $M$ and $L$, respectively. Here, $M$ represents the number of memory basis function to describe the environmental correlation functions and $L$ denotes the truncation tier. RDO and ADO are $N$-dimensional square matrix with $N$ being the size of system Fock space. These parameters, $M$, $L$ and $N$, together determine the computational cost of the HEOM calculations.

OQSs with an arbitrary Hamiltonian, including multi-impurity quantum systems.

In a schematic representation of HEOM, the hierarchy has a two-dimensional structure; see Figure 1. The vertical dimension (height) resolves the many-body correlations, while the horizontal dimension (width) resolves the non-Markovian environmental memory. RDO as well as each auxiliary density operator (ADO) is represented by an $N$-dimensional square matrix, where $N$ corresponds to the number of states spanning the system Fock space. In the case of a generic many-body OQS comprising $N_v$ impurities (orbitals), the dimension of the system Fock space is $N = (2N_s)^{N_v}$ with $N_s$ denoting the spin degrees of freedom.

In the past decade, several computer software packages have incorporated parallel programming and high-performance computing techniques, which results in the substantial acceleration of HEOM computations. This includes the graphic processing unit (GPU)-based HEOM programs[147,148] and those based on shared and distributed memory techniques[149-151]. Although these packages have demonstrated to be successful, they primarily focus on solving the bosonic HEOM. Some open-source packages, such Python-based QuTiP[152-154] and Julia-based HierarchicalEOM.jl[155], have presented numerical HEOM libraries for both bosonic and fermionic many-body OQS. Benchmark tests on the single-impurity Anderson models have exhibited the high numerical accuracy of these packages, and their applications have been extended to two-impurity Anderson models recently[156].

Zheng and collaborators have developed HEOM-QUICK (version 1), a Fortran-based numerical tool designed for investigating generic fermionic many-body OQSs[157]. HEOM-QUICK has proven its versatility in exploring a diverse range of static and dynamic properties in various many-body OQS, including quantum dots[158-160] and molecular junctions[161-162], etc. Furthermore, it has been effectively utilized in conjunction with first-principles approaches, such as the density functional theory (DFT), to examine the correlated electronic structure of realistic many-body OQSs[140-145,163-165]. The pertinent studies have been addressed in previous reviews[166,167].

To establish a closed HEOM for practical calculations, the hierarchy should be truncated in the vertical dimension at a terminal tier $L$, which is achieved by adopting deliberately designed truncation schemes in which the $L$th-tier ADOs are carefully treated. The horizontal dimension is truncated by employing various spectrum decomposition schemes to unravel hybridization correlation functions of the environment into a summation of $M$ exponential basis functions. Nevertheless, the finite dimensions of hierarchy inevitably lead to the truncation errors which potentially impair the numerical precision of the HEOM calculations. In certain scenarios, for example, when the system exhibits weak many-body correlations or when it is coupled to a high-temperature environment with strong Markovian effects, employing a low truncation depth ($L$ and $M$) is sufficient to significantly suppress the truncation errors, while ensuring the qualitative accuracy of HEOM calculations. However, in the scenario of low-temperature environments, the truncation depth of hierarchy has a profound influence on the magnitude of truncation errors and the numerical precision of HEOM calculations. This is primarily attributed to the pronounced non-Markovian memory effects and the emergence of strong many-body correlation phenomena at low temperatures. For instance, consider a quantum impurity system coupled to an electron reservoir. The Kondo effect emerges at the impurity-reservoir interface with the local spin moment of impurity screened by the conducting electrons in the environment, when the background temperature is below a characteristic temperature $T_\text{K}$ (known as the Kondo temperature)[168-170].

A low truncation depth may lead to significant truncation errors and compromises the precision of HEOM calculations. The former poses a theoretical challenge in the exploration of intricate non-Markovian dissipative dynamics and the accurately characterization of higher-order multi-electron processes. In practice, the compromised precision may also introduce considerable numerical uncertainty when solving for stationary states, thereby limiting the ability of the HEOM program to effectively distinguish between system states with small energy differences. In some cases, it may even cause numerical instability, leading to divergence in long time dynamic simulations[94,171-173].

Constructing a sufficiently large truncated hierarchy can effectively reduce the truncation errors in principle. However, HEOM calculations encounter the notorious "exponential wall" problem where the computational cost increases exponentially with $L$, $M$ and $N$. In practice, it is numerically challenging to efficiently and accurately solve the stationary state or time propagation of a many-body OQS described by an oversized hierarchy generated by large $L$ or $M$. We exemplify this challenge with the theoretical characterization of low-energy spin excitations in nanomagnets, as depicted in Figure 2. Recently, technological advancements in scanning tunnelling microscope (STM), such as the spin-polarized STM (SP-STM)[174,175] and its combination with the electron spin resonance (ESR) spectroscopy technique[176-179], have enabled measuring spin excitation energies $\Delta E$ in nanomagnets with an unprecedentedly high energy resolution (at the sub-meV or even sub-$\mu$eV level)[180-185].

The schematic diagram of the STM setup is illustrated in Figure 2(a). In this model system, a nanoscale magnet (such as a magnetic atom or molecule) with localized spin-unpaired electrons is adsorbed on the substrate. To activate spin excitations on the nanomagnet, an external energy source such as a bias voltage $V$ is required to overcome the energy gap ($\Delta E$) between different local spin states, as shown in Figure 2(d). Therefore, spin excitations often give rise to an inelastic electron tunnelling signature in the measured $I - V$ characteristic, as manifested by the changing of slope emerging under a finite bias voltage $V$ corresponding to the excitation energy $\Delta E$.

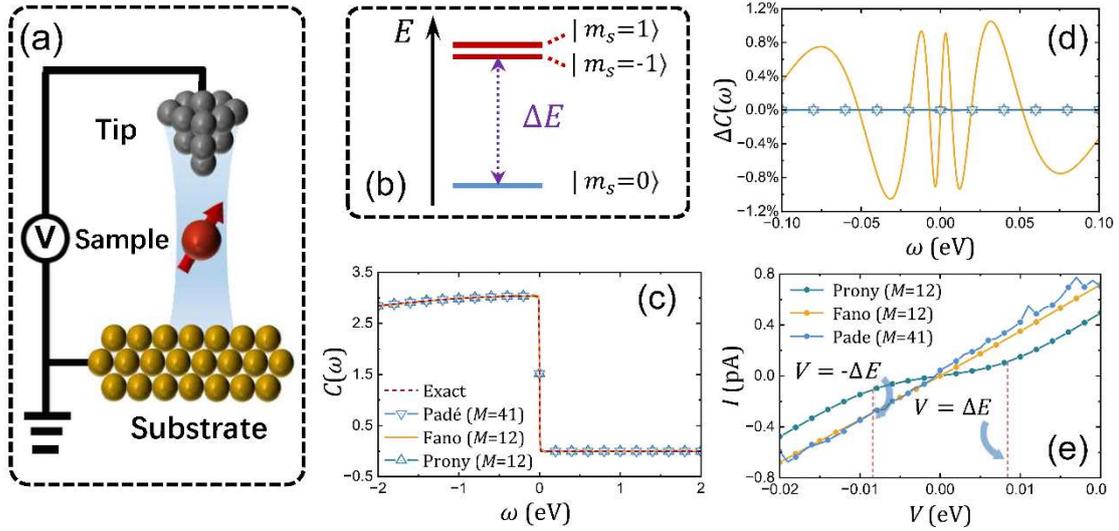

Figure 2: (a) Schematic representation of the experimental measurement on a nanomagnet embedded in an STM setup. A bias voltage is applied across the STM tip (grey) and the substrate (yellow). The red arrow represents the local spin moment of the nanomagnet whose energy diagram of the local spin states are shown in panel (b). The triplet states $|m_s\rangle, (m_s = 0, \pm 1)$ are split by the magnetic anisotropy energy $\Delta E$. Panels (c) and (d) compare the environment hybridization correlation function $C(\omega)$ and the relative error $\Delta C \equiv (C - C_{\text{Exact}})/\max\{C_{\text{Exact}}\}$ given by the Padé[186], Fano[187,188] and Prony fitting[189] spectrum decomposition schemes, respectively. The number of poles used in each decomposition scheme is given in the legend. The environment temperature is $T = 0.002$ eV. (e) The simulated $I - V$ characteristic curve. The change of slope at $V = \pm|\Delta E|$ (demonstrated by red vertical dashed lines) manifests inelastic electron tunnelling signals arising from spin excitations.

To precisely characterize the low-energy excitations, it is crucial for the numerical HEOM method to accurately reproduce the low-energy behaviour of environmental correlation functions $C(\omega) \equiv J(\omega)f(\omega)$ that captures the influence of Gaussian environments on the many-body OQS under study. Here, $J(\omega)$ and $f(\omega)$ denote the impurity-reservoir hybridization function and the Fermi distribution function, respectively, whose details will be introduced in Sec.2.1. The Padé spectrum decomposition scheme employed in HEOM-QUICK (version 1) allows for an accurate decomposition of correlation function with only a few basis functions for relatively high temperatures[186]. However, a sufficiently low temperature (e.g. the liquid Helium temperature of 4.2 K) is needed in experiments to clearly resolve low-energy spin excitations, i.e., $k_B T \ll \Delta E$. This calls for a large $M = 41$ for the Padé scheme to accurately resolve the step-like feature of $C(\omega)$ in the vicinity of Fermi level, as presented in Figure 2(c) and (d). Nevertheless, even with such a large $M$ it is still difficult to obtain an accurate $I - V$ characteristic, as manifested by the oscillatory line in Figure 2(e). Consequently, in previous works, both the temperature and the spin excitation energy had to be scaled up manually to make the calculation tractable[140,162,190]. To avoid such scaling procedure, it is essential to develop more accurate and efficient algorithms to enhance the numerical performance in the treatment of strong many-body correlation and non-Markovian memory "kernels".

Since the release of HEOM-QUICK (version 1), our focus has been on improving the efficiency and accuracy of the HEOM method. We have developed advanced HEOM methods and the corresponding numerical algorithms that are specifically designed to accurately and comprehensively characterize the many-body correlation effects and non-Markovian memory. The integration of these advancements with the previous program gives rise to the latest fermionic HEOM simulator, HEOM-QUICK2. The HEOM-QUICK2 has the following important features:

1) Inherits the strengths of HEOM-QUICK (version 1): Enable exploring a variety of many-body OQSs in the existence of diversified external fields with arbitrary time dependence (such as magnetic field, gate voltage, bias voltage, and temperature gradient); Evaluate a variety of local observables and response properties for both equilibrium and nonequilibrium scenarios[191]; Support the user-defined system models and external fields.

2) Efficiently and precisely unravels the pronounced non-Markovian memory: New Fano and Prony spectrum decomposition schemes demonstrate superior numerical performance in low-temperature environments[172,187-189], compared with the Padé spectrum decomposition scheme utilized in HEOM-QUICK (version 1). This advancement allows for the exploration of many-body OQSs coupled to much lower-temperature environments.

3) Significantly reduces the computational time and costs for solving stationary states and enhances numerical stabilities for long-time dynamics simulations of strongly correlated many-body OQSs.

4) Extends the applicability of fermionic HEOM method: Enable calculating the time-dependent response properties of many-body OQSs; Significantly enhances the "energy resolution" for low-energy excitations to the sub-meV ($< 1$ meV) level[144,192]; Combined with quantum chemistry software, HEOM-QUICK2 can precisely reproduce low-energy spin-flip excitation signatures and spin relaxation dynamics for realistic single atom/molecule junctions experimentally measured in the SP-STM setup; Explores quantum thermodynamics and thermoelectric transport in model systems[107,193-194].

This paper reviews the technical advances and new features of the HEOM-QUICK2 program, along with some examples of representative applications. Emphasis will be put on the practicality of HEOM-QUICK2 and its technical realization. The remainder of this paper is organized as follows. In Sec. 2, we introduce the quantum impurity models which HEOM-QUICK2 deals with and briefly review the fermionic HEOM method and the recent improvements. The code architecture and interfaces of HEOM-QUICK2 are present in Sec. 3, and important techniques will be exemplified by several representative applications in Sec. 4.

## 2. METHODOLOGICAL ADVANCES

### 2.1 Generic Models for Many-Body Open Quantum Systems

Consider a molecular or atomic device (system) which is coupled to macroscopic electrodes (environment). The total system is described by a generic many-body OQS as follows (we set $\hbar = k_\mathrm{B} = e = 1$ hereafter)

$$\widehat{H}_\mathrm{T}(t) = \widehat{H}_\mathrm{S}(t) + \widehat{H}_\mathrm{B} + \widehat{H}_\mathrm{hyb}. \tag{2}$$

In the second quantization formulation, by utilizing the system's electron annihilation (creation) operator $\hat{a}_{vs}(\hat{a}_{vs}^\dagger)$ which annihilates (creates) a spin-$s$ electron on the $v$th impurity, the general form of the system Hamiltonian $\widehat{H}_\mathrm{S}(t)$ with arbitrary time dependence can be constructed explicitly. This comprehensive formulation provides a complete description encompassing various types of inter- or intra-site electron-electron and spin-spin interactions, such as Coulomb repulsion, Heisenberg spin-exchange interaction, dipolar spin-spin interaction, magnetic anisotropy induced by spin-orbital coupling, etc. In Sec. 4.1, we will illustrate the construction of various interaction terms in the system Hamiltonian, using the quantum impurity model as an example.

The macroscopic electrodes are described by noninteracting electron reservoirs, i.e., $\widehat{H}_\mathrm{B} = \sum_\alpha \hat{h}_\alpha = \sum_{\alpha k s} \epsilon_{\alpha k s} \hat{d}_{\alpha k s}^\dagger \hat{d}_{\alpha k s}$. Here, $\hat{d}_{\alpha k s}$ ($\hat{d}_{\alpha k s}^\dagger$) is the annihilation (creation) operator for a spin-$s$ electron on the $k$th band of reservoir-$\alpha$ with energy $\epsilon_{\alpha ks}$. The system-environment coupling takes a linear form of $\hat{H}_{\text{hyb}} = \sum_{\alpha vs} \hat{F}^\dagger_{\alpha vs} \hat{a}_{vs} + \text{H.c.}$, where $\hat{F}^\dagger_{\alpha vs} = \sum_k t_{\alpha kvs} \hat{d}^\dagger_{\alpha ks}$, and $t_{\alpha kvs}$ is the spin-specific hopping integral between the $v$th impurity and the $k$th band of reservoir-$\alpha$. In the HEOM method, the electronic structure of reservoirs is characterized by the impurity-reservoir hybridization functions with a Lorentzian form of $J_{\alpha vv's}(\omega) \equiv \pi \sum_k t^*_{\alpha kvs} t_{\alpha kv's} \delta(\omega - \epsilon_{\alpha ks}) = \Gamma_{\alpha vv's} W_\alpha^2 / [(\omega - \Omega_\alpha)^2 + W_\alpha^2]$, where $\Omega_\alpha$ and $W_\alpha$ are the band center and band width of reservoir-$\alpha$, respectively; and $\Gamma_{\alpha vv's}$ constitutes an effective hybridization matrix $\mathbf{\Gamma}_{\alpha s}$ between the impurity and reservoir-$\alpha$. In particular, the spin-specific hybridization matrix $\mathbf{\Gamma}_{\alpha s}$ plays a critically important role in simulating and interpreting the differential conductance ($dI/dV$) spectra experimentally measured by the SP-STM setup[143].

Since the linearly coupled non-interacting electron reservoirs satisfy the Gaussian statistics, their influence on the reduced system dynamics is fully captured by the environmental correlation functions, which are related to the hybridization functions via the fluctuation-dissipation theorem for the fermionic grand canonical ensembles:

$$C^\sigma_{\alpha vv's} \equiv \langle \hat{F}^\sigma_{\alpha vs}(t) \hat{F}^{\bar{\sigma}}_{\alpha vs}(\tau) \rangle_{\text{B}} = \frac{1}{\pi} \int_{-\infty}^{+\infty} d\omega\, J^\sigma_{\alpha vv's}(\omega)\, f^\sigma_\alpha(\omega)\, e^{i\sigma\omega t}. \tag{3}$$

Here, $\sigma = \pm$ and $\bar{\sigma} = -\sigma$; $\hat{F}^+_{\alpha vs}(t) \equiv e^{i\hat{H}_{\text{B}} t} \hat{F}^\dagger_{\alpha vs} e^{-i\hat{H}_{\text{B}} t}$ and $\hat{F}^-_{\alpha vs}(t) \equiv e^{i\hat{H}_{\text{B}} t} \hat{F}_{\alpha vs} e^{-i\hat{H}_{\text{B}} t}$; $J^+_{\alpha v'vs}(\omega) = J^-_{\alpha vv's}(\omega) = J_{\alpha vv's}(\omega)$; and $f^\sigma_\alpha(\omega) = 1/[1 + e^{\sigma \beta_\alpha (\omega - \mu_\alpha)}]$ is the Fermi distribution function, with $\beta_\alpha \equiv 1/T_\alpha$ and $\mu_\alpha$ being the inverse temperature and the chemical potential of reservoir-$\alpha$, respectively. The expectation value is evaluated by averaging over the reservoir's degrees of freedom, i.e., $\langle \hat{O} \rangle_{\text{B}} \equiv \text{tr}_{\text{B}}[\hat{O} \rho^{\text{eq}}_B]$, where $\rho^{\text{eq}}_B \equiv \prod_\alpha e^{\sigma \beta_\alpha (\hat{h}_\alpha - \mu_\alpha \hat{N}_\alpha)} / \text{tr}_{\text{B}}[e^{\sigma \beta_\alpha (\hat{h}_\alpha - \mu_\alpha \hat{N}_\alpha)}]$ is the equilibrium density matrix of isolated reservoirs and $\hat{N}_\alpha \equiv \sum_{k,s} \hat{d}^\dagger_{\alpha ks} \hat{d}_{\alpha ks}$ is the electron number operator of reservoir-$\alpha$. Moreover, the correlation function effectively characterizes the complete influence of external fields (including bias voltage, temperature gradient, etc) on reservoirs. For instance, consider a time-dependent bias voltage $V_{\alpha s}(t)$ which shifts the chemical potential $-eV_{\alpha s}(t) \equiv \mu_{\alpha s}(t) = \mu^{\text{eq}}_{\alpha s} + \Delta\mu_{\alpha s}(t)$ of spin-$s$ electrons of reservoir-$\alpha$. Here, $\mu^{\text{eq}}_{\alpha s}$ is the equilibrium chemical potential (generally set $\mu^{\text{eq}}_{\alpha s} = 0$) and $\Delta\mu_{\alpha s}(t)$ accounts for the time-dependent components, respectively. The presence of $\Delta\mu_{\alpha s}(t)$ introduces an extra phase factor on the environmental correlation function as follows

$$\tilde{C}^\sigma_{\alpha v\,'s}(t,\tau) = \exp\left[\sigma i \int_\tau^t dt'\, \Delta\mu_{\alpha s}(t)\right] C^\sigma_{\alpha v\,'s}(t - \tau). \tag{4}$$

## 2.2 Fermionic HEOM Method

For a generic fermionic many-body OQS coupled to a collection of noninteracting electron reservoirs, the fermionic HEOM is[80,195]

$$\dot{\rho}^{(n)}_{j_1\ldots j_n} = \left(-i\mathcal{L}_S + \sum_r \gamma_{j_r}\right)\rho^{(n)}_{j_1\ldots j_n}$$

$$-i\sum_j [\hat{a}^{\bar{\sigma}}_{vs}\rho^{(n+1)}_{jj_1\ldots j_n} - (-1)^n \rho^{(n+1)}_{jj_1\ldots j_n}\hat{a}^{\bar{\sigma}}_{vs}]$$

$$-i\sum_{r,v'}[(-1)^{r-1}\eta_{j_r v'}\hat{a}^{\sigma}_{v's}\rho^{(n-1)}_{j_1\ldots j_{r-1}j_{r+1}\ldots j_n} - (-1)^{n-r}\eta^*_{\bar{j}_r v'}\rho^{(n-1)}_{j_1\ldots j_{r-1}j_{r+1}\ldots j_n}\hat{a}^{\sigma}_{v's}]. \quad (5)$$

Here, we introduce a multicomponent index $j = \{\sigma\alpha vps\}$ to label a principal dissipation mode, which corresponds to the transfer of a spin-$s$ electron to (from) the $v$th impurity of the system from (to) reservoir-$\alpha$ associated with the characteristic dissipation rate $\gamma_j$. The conjugate index is denoted as $\bar{j} = \{\bar{\sigma}\alpha vps\}$ and conforms to the relation of $\gamma_{\bar{j}} = \gamma_j^*$. The complex coefficients $\eta_{jv'} = \eta^{\sigma}_{\alpha vv'ps}$ arise from the spectral decomposition of reservoir correlation functions; see Equation (6).

To construct the formally closed HEOM in the horizontal dimension, the reservoir correlation function is unravelled by a series of exponential functions as[80,186]

$$C^{\sigma}_{\alpha vv's}(t-\tau) \approx \sum_{p=1}^P C^{\sigma}_{\alpha vv'ps}(t-\tau) = \sum_{p=1}^P C_{jv'}(t-\tau) = \sum_{p=1}^P \eta_j\, e^{-\gamma_j(t-\tau)}. \quad (6)$$

Here, $M = P$ is the number of exponential basis functions to unravel the reservoir memory. The decomposition of correlation functions in Equation (6) can be realized by employing the sum-over-poles expansion technique and applying Cauchy's residue theorem to the numerical evaluation of Equation (3). At high temperatures, the Matsubara[80] and the Padé[186,196] spectrum decomposition schemes demonstrate high accuracy and efficiency using a few exponential functions, However, these methods become increasingly expensive at low temperatures, as they require a rather large $P$ to achieve a quantitatively accurate description of the correlation function.

To close the hierarchy in a finite vertical dimension, several truncation schemes (or terminators) have been developed. These includes the zero-value truncation scheme, which sets $\rho^{(L)}_{j_1\ldots j_L} = 0$ at the $L$th tier, and the time-derivative truncation scheme $\dot{\rho}^{(L)}_{j_1 \cdots j_L} = 0$ at $L$th-tier[197]. These truncation schemes yield accurate results, provided that numerical convergence with respect to the truncation tier $L$ is reached. However, these two schemes exhibit slow convergence of the results with respect to $L$ in the treatment of strongly correlated OQSs[198], and occasionally encounter numerical instabilities in simulating long-time dissipative dynamics[172]. In the following subsections, we will review recently developed numerical algorithms which effectively reduce the truncation errors.

**2.3 Handling Truncation Errors**

The key to reducing the truncation error in the horizontal dimension is to develop a spectrum decomposition scheme that can precisely reproduce the low-energy feature of reservoir correlation functions for low-temperature environment by as few basis functions as possible. Cui *et al*. have proposed the Fano spectrum decomposition scheme which effectively reduces the residual errors in the low-energy regime by introducing a few generalized Fano functions as a low-temperature improvement to the Padé scheme[187]. Specifically, the Fano scheme defines a higher reference temperature $T_0$ by scaling up the original environmental temperature $T$ by a factor of $\chi$. The Padé expansion of Fermi function is then performed at the reference temperature $T_0$. A few Padé poles generally lead to a reasonably accurate decomposition as long as $\chi$ is sufficiently large. The residual error between the approximation and exact curves exhibits asymmetric Fano-like lineshape. The generalized Fano functions are adopted as a low-temperature correction to the Padé expansion, and the parameters in the Fano functions are determined by minimizing the residual error. Finally, applying the sum-over-pole technique yields the decomposed coefficients $\{\eta_j\}$ and dissipation rates $\{\gamma_j\}$ for the generalized Fano functions. The reservoir correlation function is unravelled by

$$C(t - \tau) = C^{\text{Padé}}(t - \tau) + C^{\text{Fano}}(t - \tau)$$

$$= \sum_{p=1}^{P} \eta_p \, e^{-\gamma_p (t-\tau)} + \sum_{q=1}^{Q} \eta_q (t - \tau)^{m_q} \, e^{-\gamma_q (t-\tau)} . \quad (7)$$

Here, $Q$ is the number of polynomial exponential functions given by the Fourier transform of the generalized Fano functions and the total number of dissipation modes in the Fano scheme is determined by $M = P + Q$. All the superscripts and subscripts of $C(t)$ are omitted for brevity. The polynomial terms improve the performance of decomposition over the Padé scheme in describing the slow-decay of $C(t)$ in the long-time regime.

The numerical benchmarks also demonstrate the superior efficiency of the Fano scheme over the Padé scheme to unravel the memory "kernel" for low-temperature environment[188]. Employing the Fano-based HEOM method, Zhang *et al*. have explored the Kondo correlation features and time-dependent charge transport of quantum impurity systems embedded in low temperature environment[188]. Zhuang *et al*. have simulated the asymmetric splitting of Kondo peak of a Co atom deposited on a Cu(100) substrate in a spin-polarized STM setup[142].

Recently, Chen *et al*. have proposed the time-domain Prony spectral fitting decomposition, which employs a series of exponential functions as the basis functions to unravel the real and imaginary parts of the reservoir correlation function $C(t)$, respectively, as follows[189]

$$C(t-\tau) = \text{Re}[C(t-\tau)] + i\text{Im}[C(t-\tau)]$$
$$= \sum_{p=1}^{P} \xi_p\, e^{-\gamma_p(t-\tau)} + i \sum_{q=1}^{Q} \zeta_q\, e^{-\lambda_q(t-\tau)}. \quad (8)$$

The total number of exponential functions used is $M = P + Q$. For electron reservoirs with a Lorentzian-type hybridization function, the real part of correlation function $C(t)$ is a single exponential function, i.e., $\text{Re}[C(t)] = (\Gamma W/2)\exp(-Wt)$; while its imaginary part can be evaluated with reference to the Prony fitting protocol as detailed in Ref. [189].

Figure 3 compares the numerical performance of different spectrum decomposition schemes to unravel reservoir correlation function $C(t)$ and examine the relative deviation $\delta C(\omega) \equiv [C_{\text{dec}}(\omega) - C_{\text{Exact}}(\omega)]/\max\{C_{\text{Exact}}(\omega)\}$ at different temperatures $T$. Here, $C_{\text{Exact}}(\omega)$ is the exact frequency-domain correlation function, while $C_{\text{dec}}(\omega)$ is an approximation calculated by a decomposition scheme. The reservoir hybridization function assumes a Lorentzian form. In a case of $T/\Gamma = 0.35$, the Padé scheme with 6 poles exhibits a large deviation from the exact value even in the low-frequency region; see the inset of Figure 3(a). This deviation diminishes significantly as the number of poles increases. In contrast, the Fano scheme involves 2 poles given by the Padé expansion and 10 poles from the generalized Fano functions whose parameters are listed in Ref. [188], and the resulting $C(\omega)$ agrees reasonably well with the exact curve, but still presents appreciable errors in the low-frequency region, as shown in Figure 3(b). It is remarkable that the Prony scheme precisely reproduces the exact curve with only 6 poles and yields a vanishingly small deviation in Figure 3(b). Even for a very low reservoir temperature of $T/\Gamma = 0.01$, the Prony scheme still yields accurate $C(\omega)$ with 9 poles in Figure 3(c), and results in the significantly reduced $\delta C(\omega)$ distributed tightly around zero in Figure 3(d). In contrast, the Padé scheme requires many more poles to attain reasonably accurate $C(\omega)$, and the resulting HEOM calculations inevitably become more expensive. The difficulty of treating low temperatures by the Fano scheme becomes apparent by examining the low-frequency region. While the obtained $C(\omega)$ remains accurate in the high-frequency region, pronounced oscillations arise in the low-frequency region, as illustrated in Figure 3(d).

We now revisit the numerical benchmarks in Figure 2(b-e) and examine the performance of the Fano and Prony schemes for exploring low-energy spin excitations. As depicted in Figure 2(e), due to the superior accuracy in reproducing the exact $C(\omega)$, the Prony scheme using 12 poles correctly captures the inelastic electron tunnelling signatures upon the occurrence of spin excitation, as manifested by the change of slope emerging when the bias voltage matches the excitation energy $V = \pm|\Delta E|$. In contrast, although the Fano scheme with 15 basis functions overall reproduces the exact correlation function, the minor residual error in the low-frequency region (represented by the oscillatory line in Figure 2(d)) still affects the calculated system properties. Consequently, the resulting $I - V$ characteristics fails to reproduce the inelastic tunnelling features related to the spin excitations at $V = \pm|\Delta E|$. Despite the compromised performance on low-energy spin excitations, it is emphasized that the Fano scheme allows for quantitative characterization of the strength of Kondo correlation in quantum impurity models at low temperatures, as demonstrated later in Figure 4.

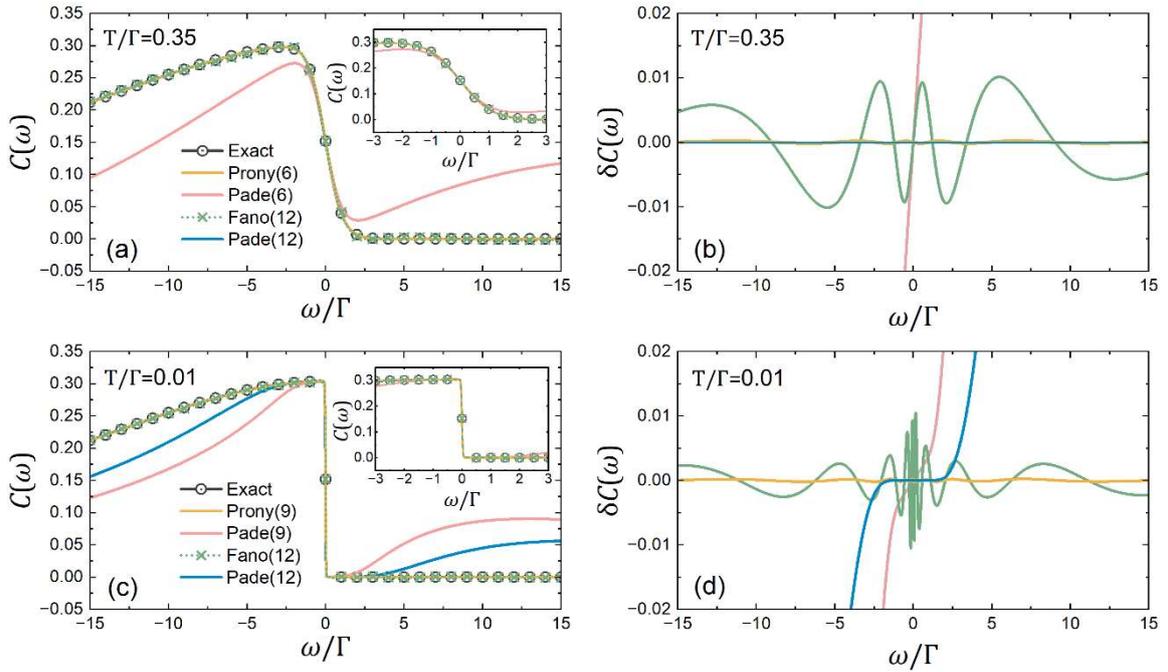

Figure 3: (a) Reservoirs correlation functions $C(\omega)$ and (b) the relative deviations $\delta C(\omega) \equiv [C_{\text{dec}}(\omega) - C_{\text{Exact}}(\omega)]/\max\{C_{\text{Exact}}(\omega)\}$ obtained by the Padé, Fano, and Prony spectrum decomposition schemes (coloured lines) at a high temperature $T/\Gamma = 0.35$. Here, $C_{\text{Exact}}(\omega)$ is the exact frequency-domain correlation function, while $C_{\text{dec}}(\omega)$ is an approximation calculated by a decomposition scheme. The number of poles (basis functions) used in each spectrum decomposition scheme is given in the legend. Panels (c) and (d) show the correlation functions $C(\omega)$ and relative deviation $\delta C(\omega)$ at a low temperature $T/\Gamma = 0.01$, respectively. The insets magnify the low-frequency region. The band width of reservoirs with a Lorentzian-type hybridization function is $W/\Gamma = 15$.

Several time-domain fitting schemes have been developed as alternatives to the Prony method. These include the matrix-pencil method[199], the estimation of signal parameters via rotational invariance techniques (ESPRIT)[200], and the estimation of signal parameters based on iterative rational approximation (ESPIRA)[201]. Compared to frequency-domain approaches like the Matsubara, Padé, and Fano schemes, these time-domain methods can more accurately decompose hybridization correlation functions using fewer poles[202]. However, they require a sufficiently long cutoff time to capture long-time dynamic behaviour accurately. Additionally, the time-domain schemes generally take more time and memory to fit correlation functions than frequency-domain approaches.

We have already performed the Prony fitting spectrum decomposition at various temperatures and stored the resulting parameters in the HEOM-QUICK2 presets, making it readily accessible for users. The Prony scheme employed here adopts a cut-off time much longer than the inverse of temperature, i.e. $t_{\text{cut}} \gg 1/T$, which exceeds the ending time of most dissipative dynamics. Moreover, HEOM-QUICK2 also provides the Fortran and Python codes of the Prony scheme for any temperature, enabling users to unravel $C(t)$ as needed.

Handling the truncation error in the vertical dimension is more challenging and less explored than the spectrum decomposition in the horizontal dimension. For generic quantum impurity systems, Han *et al.* have analytically demonstrated that the $(n + \bar{L})$th-tier zero-value truncation scheme with $\bar{L} = 2N_\sigma N_s N_v$ yields the numerically exact ADO from 0th-tier up to $n$th-tier[195]. $N_\sigma = 2$ for fermionic environments and $N_s$ and $N_v$ denote the spin and impurity (orbital) degrees of freedom, respectively.

Although $\bar{L}$ could be extremely large in practical HEOM calculations, numerous works have reported that a truncation tier $L$ lower than $\bar{L}$ still gives rise to numerically accurate results[203-206]. In practical HEOM calculations, the depth of truncation tier $L$ generally depends on the strength of many-body correlations[198]. For instance, the zero-value terminator at a low truncation tier $L$ enables yielding sufficiently precise results in a case of weak many-body correlation, while for strongly correlated OQSs a deep $L$ is necessary to obtain quantitatively accurate results.

Recently, we have developed an adiabatic terminator by decoupling the fastest principal dissipation mode from the slower ones for the $L$th-tier ADO[172]

$$\rho^{(L)}_{j_1\ldots j_L} = -i \sum_{\nu,s} \left[ \mathcal{W}_{j_r\nu s}\, \hat{a}^{\sigma_r}_{\nu s}\, \rho^{(L-1)}_{j_1\ldots j_{r-1}j_{r+1}\ldots j_L} + \mathcal{W}^\dagger_{j_r\nu s}\, \rho^{(L-1)}_{j_1\ldots j_{r-1}j_{r+1}\ldots j_L}\, \hat{a}^{\sigma_r}_{\nu s} \right]. \qquad (9)$$

Here, the reduced Liouville propagator $\mathcal{W}_{j_r\nu s}$ explicitly accounts for the exchange of a spin-$s$ electron between the $\nu$th impurity and environment via the $j_r$th dissipation mode with $\hat{a}^{-}_{\nu s} = \hat{a}_{\nu s}$ and $\hat{a}^{+}_{\nu s} = \hat{a}^\dagger_{\nu s}$.

By employing the above state-of-the-art methods, HEOM-QUICK2 handles strongly correlated OQSs more efficiently than its predecessors. To demonstrate this, we carry out comprehensive benchmark tests on the widely studied single impurity Anderson model (SIAM) and examine the convergence of different system properties with respect to the truncation tier $L$. The impurity is represented by the Hamiltonian

$$\hat{H}_S = \sum_s \epsilon_s \hat{n}_s + U \hat{n}_\uparrow \hat{n}_\downarrow.$$

In the following, we adopt a large Coulomb interaction energy $U = 10\,\Gamma$ and a low temperature $T = 0.015\,\Gamma$. The Kondo temperature evaluated by[170] $T_K = \sqrt{\Gamma U/2}\, e^{-\pi U/8\Gamma + \pi\Gamma/2U} \approx 0.05\,\Gamma$ is higher than $T$, thus leading to a prominent Kondo resonance.

We first employ the HEOM-QUICK2 program to revisit the equilibrium SIAM coupled to a single reservoir, which was tested by HEOM-QUICK (version 1) in the previous review[157]. HEOM-QUICK2 employs the adiabatic terminator and HEOM-QUICK (version 1) adopts the zero-value scheme. HEOM-QUICK2 applies the Prony fitting scheme with 12 basis functions to unravel low-temperature reservoir correlation functions, whereas the Padé scheme used in HEOM-QUICK (version 1) requires 34 basis functions to reach the same decomposition accuracy.

Table 1 demonstrates the convergence of diagonal elements of RDO $\rho_{\uparrow,\downarrow}$ and internal energy $E_{\text{sys}}$

Table 1: The numerical benchmark on a SIAM connected to a single reservoir by using (a) the HEOM-QUICK2 and (b) the HEOM-QUICK (version 1) programs, respectively.

(a) Results of HEOM-QUICK2 with the Prony scheme and the adiabatic terminator

| $L$ | $\rho_\uparrow = \rho_\downarrow$ | $E_{\text{sys}}$ | Number of ADOs | Physical memory (MB) | CPU time (s) |
|---|---|---|---|---|---|
| 1 | 0.425919 | -0.851839 | 25 | 0.02 | 7 |
| 2 | 0.437410 | -0.874821 | 889 | 0.69 | 65 |
| 3 | 0.433215 | -0.866430 | 18,169 | 8.53 | 606 |
| 4 | 0.431632 | -0.863264 | 412,153 | 131 | 5,419 |
| 5 | 0.431795 | -0.863590 | 7,379,449 | 1,233 | 54,824 |
| 6 | 0.431664 | -0.863328 | 138,762,745 | 14,000 | 538,479 |

(b) Results of HEOM-QUICK (version 1) with the Padé scheme and the zero-value terminator

| $L$ | $\rho_\uparrow = \rho_\downarrow$ | $E_{\text{sys}}$ | Number of ADOs | Physical memory (MB) | CPU time (s) |
|---|---|---|---|---|---|
| 1 | 0.500000 | -0.999999 | 71 | 0.03 | 10 |
| 2 | 0.455507 | -0.911013 | 7,421 | 2.39 | 30 |
| 3 | 0.456971 | -0.913943 | 436,171 | 69.76 | 1,593 |
| 4 | 0.435365 | -0.870729 | 28,948,046 | 3,393 | 34,103 |
| 5* | N/A | N/A | 1,499,560,546 | N/A | N/A |

Table 1: The numerical benchmark on a SIAM connected to a single reservoir by using (a) the HEOM-QUICK2 and (b) the HEOM-QUICK (version 1) programs, respectively. The benchmark was done on a workstation with Intel(R) Xeon(R) Gold 6248R CPU @ 3.00GHz with 64 GB memory. 24 CPUs were used for the parallel solution to HEOM equations via the TFQMR algorithm, and the convergence criterion is the same for two programs. Here lists the truncation tier $L$, diagonal elements of RDO $\rho_{\uparrow,\downarrow}$, internal energy $E_{\text{sys}}$, number of ADOs, physical memory (in unit of MB), and the CPU time (in unit of second) required for calculating an equilibrium state of SIAM. Marked by the asterisk, the HEOM-QUICK (version 1) calculation with $L = 5$ is not available because the memory required by the tremendous number of ADOs is out of memory or system resources outlook. The energetic parameters adopted here are (in the unit of $\Gamma$): $\epsilon_\uparrow = \epsilon_\downarrow = -U/2 = -5$, $T = 0.015$ and $W = 15$.

with respect to the truncation tier $L$ and compares the computational costs required for each HEOM calculation. Compared with the results obtained by HEOM-QUICK (version 1), HEOM-QUICK2 shows a more rapid and uniform convergence behaviour as $L$ increases, while requiring less physical memory and CPU time at a specific truncation tier. For instance, in the case of $L = 4$, HEOM-QUICK2 exhibits a remarkably low discrepancy of 0.007% for the values of $\rho_{\uparrow,\downarrow}$ and $E_{\text{sys}}$, compared to the results obtained at $L = 6$; see Table 1(a). In contrast, HEOM-QUICK (version 1) shows a relatively high discrepancy of 0.9% for these properties; see Table 1(b). Moreover, it is observed that the calculations at $L = 5$ is already too demanding for HEOM-QUICK (version 1) because of the large number of basis functions required by the Padé scheme, which leads to a drastic growth in the number of ADOs and the physical memory required to store all the nonzero elements of these ADOs as $L$ increases.

(a) Results of HEOM-QUICK2 with the Prony scheme and the adiabatic terminator

| $L$ | $\rho_\uparrow = \rho_\downarrow$ | $I$ (pA) | Number of ADOs | Physical memory (MB) | CPU time (s) |
|---|---|---|---|---|---|
| 1 | 0.542137 | -4323.34 | 49 | 0.07 | 15 |
| 2 | 0.379988 | 51.28 | 2,929 | 3.85 | 212 |
| 3 | 0.375376 | 71.12 | 106,609 | 109.31 | 3,358 |
| 4 | 0.371751 | 76.49 | 3,631,729 | 2,890 | 89,964 |
| 5 | 0.371625 | 76.56 | 102,169,201 | 56,485 | 1,194,708 |

(b) Results of HEOM-QUICK (version 1) with the Padé scheme and the zero-value terminator

| $L$ | $\rho_\uparrow = \rho_\downarrow$ | $I$ (pA) | Number of ADOs | Physical memory (MB) | CPU time (s) |
|---|---|---|---|---|---|
| 1 | 0.500000 | 5.26 | 141 | 0.03 | 15 |
| 2 | 0.418053 | 25.18 | 24,641 | 7.81 | 148 |
| 3 | 0.438280 | -6816.76 | 2,597,141 | 526.98 | 183,267 |
| 4 | 0.380760 | 85.10 | 257,703,391 | 41,007 | 795,380 |
| 5 | N.A. | N.A. | 21,056,365,891* | N.A. | N.A. |

Table 2: The numerical benchmark on a SIAM subjected to a constant bias voltage by using (a) the HEOM-QUICK2 and (b) the HEOM-QUICK (version 1) programs, respectively. The benchmark was done on a workstation with Intel(R) Xeon(R) Gold 6248R CPU @ 3.00GHz with 64 GB memory. 24 CPUs were used for the parallel solution to HEOM equations via the TFQMR algorithm, and the convergence criterion is the same for two programs. Here lists the truncation tier $L$, diagonal elements of RDO $\rho_{\uparrow,\downarrow}$, electric current $I$, number of ADOs, physical memory (in unit of MB), and the CPU time (in unit of second) required for calculating a nonequilibrium stationary state of SIAM. T Marked by the asterisk, the HEOM-QUICK (version 1) calculation with $L = 5$ is not available because the memory required by the tremendous number of ADOs is out of memory or system resources outlook. The energetic parameters adopted here are (in the unit of $\Gamma$): $\epsilon_\uparrow = \epsilon_\downarrow = -U/2 = -5$, $T_L = T_R = 0.015$, $V_L = 0.005$, $V_R = 0$ and $W_L = W_R = 15$.

We then compare the numerical performance of two programs in solving non-equilibrium stationary state of SIAM subjected to a constant bias voltage $V$. As presented in Table 2(a), although there is a significant deviation in the system observables calculated by HEOM-QUICK2 at $L = 2$, these observables converge rapidly with the increasing $L$ and attain quantitative accuracy at $L = 4$. This is largely attributed to the adiabatic terminator for effectively handling the many-body correlation and the Prony scheme for efficiently unravelling the hybridization correlation functions. However, the results of HEOM-QUICK (version 1) demonstrate a substantial deviation at $L = 3$ and have not yet converged at $L = 4$, as evidenced by Table 2(b). Furthermore, HEOM-QUICK2 substantially reduces the cost of physical memory and CPU time, leading to remarkable improvement in the computational efficiency for non-equilibrium OQSs.

The convergence of other system properties has been extensively examined in previous works. Zhang *et al.* have evaluated the hybridization energy $E_{\text{hyb}}$ up to the converged tier; see Figure 4(a)[172]. Compared with the zero-value terminator, the adiabatic terminator converges more smoothly with respect to $L$. Figure 4(b) exhibits the impurity spectral function $A_s(\omega)$ of SIAM with the prominent Kondo correlation. HEOM-QUICK (version 1) at $L = 3$ does not offer a correct prediction to $A_s(\omega)$ and a higher truncation tier is required. Although HEOM-QUICK (version 1) at $L = 4$ yields the accurate Hubbard peaks resided at $\omega = \epsilon$ and $\omega = \epsilon + U$, but it apparently overestimates the strength of Kondo correlation. The value of $A_s(\omega = 0)$ quantitatively converges at $L = 5$; see the inset of Figure 4(b). HEOM-QUICK2 not only yields the converged Hubbard peaks in the vicinity of the $\omega = \epsilon$ and $\epsilon + U$ at a low tier $L = 3$, but also accurately captures the feature of Kondo resonance at the Fermi level.

Ding *et al.* have further investigated the convergence of the zero-frequency impurity spectral function $A_0 \equiv A_s(\omega = 0)$ of strongly correlated OQSs with varying system-environment coupling strengths in Figure 4(c)[198]. The curves of the adiabatic terminator at different $L$ not only vary smoothly with respect to $\Gamma$, as evidenced by the lines in Figure 4(d), but also agree reasonably well with the analytic results predicted by the Friedel sum rule. In contrast, the zero-value terminator gives rise to the unphysically oscillatory lines.

We now illustrate the enhanced numerical stability of dissipative dynamics calculations achieved by HEOM-QUICK2 with the adiabatic terminator by presenting two representative cases. Our primary focus is to examine the influence of the truncation scheme, and thus both programs employ the same

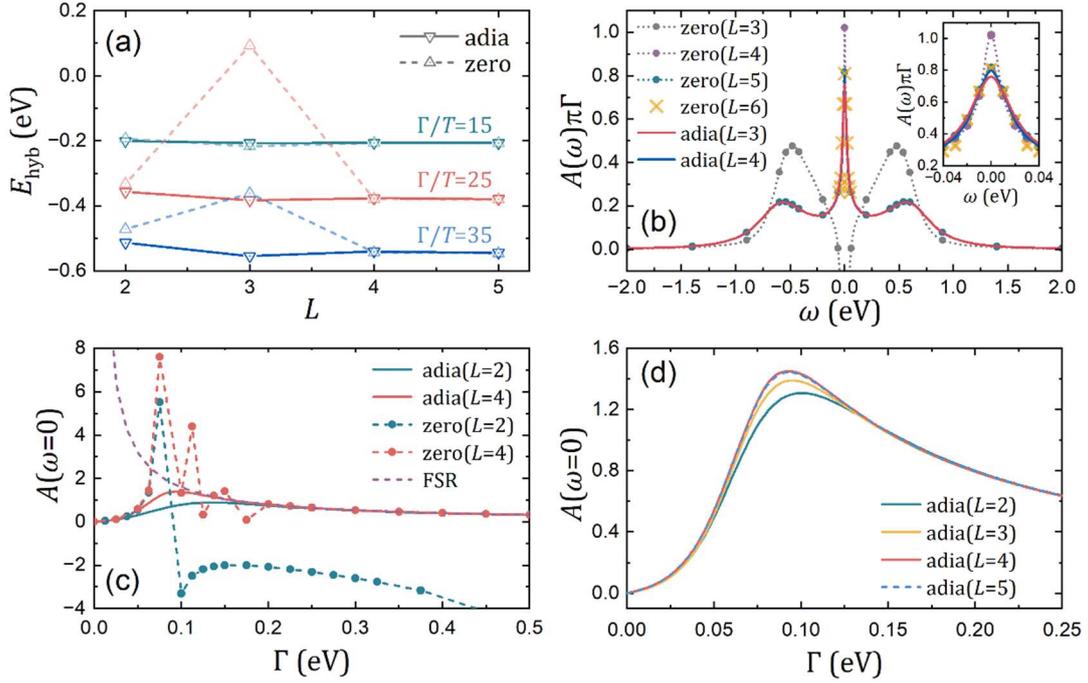

Figure 4: Benchmarks for the adiabatic terminator on quantum impurity models. (a) The hybridization energy $E_{\text{hyb}}$ and (b) impurity spectral function $A(\omega)$ calculated by the adiabatic terminators (adia) and by the zero-value terminator (zero) at different truncation tiers $L$, respectively[172] (Copyright from American Institute of Physics in 2021 with permission). The other parameters adopted in (a) are (in the unit of $U$): $\epsilon_\uparrow = \epsilon_\downarrow = -0.5$, $W = 2.5$ and $T = 0.005$. Those adopted in (b) are (in the unit of $U$): $\Gamma = 0.125$, $\epsilon_\uparrow = \epsilon_\downarrow = -0.5$, $W = 2.5$ and $T = 0.005$; (c) Zero-frequency impurity spectral function $A_0$ as a function of the system-reservoir coupling $\Gamma$ for the single impurity Anderson model[198]. The brown dashed line represents the analytic results given by the Friedel sum rule at exact zero temperature. Panel (d) magnifies the results of the adiabatic terminator in an intermediate $\Gamma$ region (Copyright from American Institute of Physics in 2022 with permission). The other energetic parameters in (c) and (d) are (in the unit of eV): $\epsilon_\uparrow = \epsilon_\downarrow = -U/2 = -1.0$, $W = 15$ and $T = 0.002$.

Fano scheme in the subsequent calculations to unravel the reservoir correlation functions of low-temperature reservoirs accurately and efficiently.

One case is the time-dependent electric current of SIAM driven by sinusoidal *ac* voltages. The SIAM is initially prepared in an equilibrium state, and sinusoidal voltages $\mu_L(t) = -\mu_R(t) = V_0 \sin(\omega_0 t)$ are applied to the left and right reservoirs after $t = 0$, where $V_0$ and $\omega_0$ are the magnitude and frequency of the *ac* voltages, respectively. The other case describes the charge transfer of a two-impurity Anderson model coupled to a single reservoir. Initially, impurity-1 has the electron-hole symmetry, i.e., $\epsilon_{1s} = -U/2$, and is thus half-filled, while the levels of impurity-2 are above the Fermi level, i.e., $\epsilon_{2s} = U/2$, and therefore are unoccupied. A gate voltage is imposed on the two impurities after $t = 0$, which effectively swaps their energy levels. The time-dependent Hamiltonian of the two impurities is represented by $\hat{H}_S = \sum_s \epsilon_{1s}(t)\,\hat{n}_{1s} + \epsilon_{2s}(t)\,\hat{n}_{2s} + \sum_\nu U\,\hat{n}_{\nu\uparrow}\hat{n}_{\nu\downarrow}$ with the energy levels being $\epsilon_{1s}(t<0) = \epsilon_{2s}(t \geq 0) = U/2$ and $\epsilon_{1s}(t \geq 0) = \epsilon_{2s}(t<0) = -U/2$.

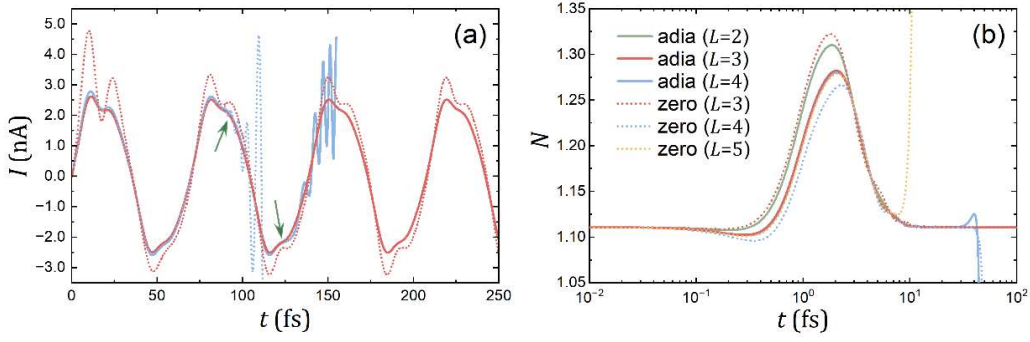

Figure 5: The numerical benchmarks for (a) time-dependent electric currents $I(t)$ of a SIAM driven by an *ac* voltage and (b) the time evolution of total electron occupancy $N(t)$ of a two-impurity Anderson model triggered by changing the impurity levels. (Copyright from American Institute of Physics in 2021 with permission) The results of the adiabatic and zero-value terminators are represented by solid and dotted lines, respectively. The green arrows in (a) mark the times from which the divergence becomes prominent. The benchmark was done on a workstation with Intel(R) Xeon(R) Gold 6248R CPU @ 3.00GHz with 64 GB memory. 24 CPUs were used for the parallel solution to stationary states via the TFQMR algorithm and the sequent dynamic simulations via the 4th-order Runge-Kutta method. The energetic parameters adopted for SIAM are (in the unit of $\Gamma$) : $\epsilon_\uparrow = \epsilon_\downarrow = -U/2 = -5, T_L = T_R = 0.015, V_L = 0.005, V_R = 0, \omega_0 = 0.12$ and $W_L = W_R = 15$, and for the two-impurity Anderson model are (in the unit of $\Gamma$): $\epsilon_{1s}(t<0) = \epsilon_{2s}(t \geq 0) = -\epsilon_{1s}(t<0) = -\epsilon_{2s}(t \geq 0) = -U/2 = 10, T_L = T_R = 0.01, V_L = V_R = 0$ and $W_L = W_R = 15$.

Figure 5(a) and (b) show the real-time response current of SIAM subjected to a sinusoidal *ac* voltage and the time evolution of total impurity occupancy $\langle \hat{N} \rangle(t) = \sum_{\nu,s}\langle \hat{n}_{\nu s}\rangle(t)$ calculated at different $L$, respectively. It is noted that the system observables calculated by HEOM-QUICK2 converge at $L = 3$ and present their numerical stability until the end of calculations in both cases. This can be attributed to the adiabatic scheme which significantly improves the convergence of system properties with respect to $L$, and thus lead to a quantitatively accurate value at a low tier where the divergence does not occur. Moreover, the adiabatic terminator slightly delays the occurrence of divergence for the case of $L = 4$, as manifested by the green arrows in Figure 5(a). This indicates that the adiabatic scheme can indeed alleviate numerical instability to some extent in specific scenarios. Although HEOM-QUICK (version 1) with the zero-value scheme achieves its numerical convergence at $L = 5$ for both $I(t)$ and $\langle \hat{N} \rangle$, it still diverges after a certain time instant before the end of dynamics.

## 2.4 System Fock Space Compression

As the truncation tier $L$ and the size of system Fock space increase, the computer memory required to save the numerous ADO matrices and the computation time grows exponentially. To tackle this problem, the sparse matrix technique proposed by Hou *et al.* has been employed to solve HEOM[197]. Based on this useful technique, we have developed the subsystem HEOM method by discarding the high-energy excited states of many-body OQSs and then projecting the HEOM onto the low-energy subsystem[192]. To achieve this, we first define the projection operator $\hat{P} = \sum_{l=1}^{\widetilde{N}} |l\rangle\langle l|$ with $\{|l\rangle\}$ denoting the low-energy eigenstates of $\hat{H}_S$. Here, the dimension of the subsystem is represented by $\widetilde{N}$. We then replace the system operators in Equation (5) by the corresponding subsystem operators, i.e., the annihilation and creation operators $\hat{a}_{vs}^{\sigma} \to \tilde{a}_{vs}^{\sigma} \equiv \hat{P}\hat{a}_{vs}^{\sigma}\hat{P}$ and the system Liouvillian $\mathcal{L}_S(\cdot) \to \widetilde{\mathcal{L}}_S(\cdot) \equiv [\hat{P}\hat{H}_S\hat{P}, (\cdot)]$.

The subsystem HEOM method is applied to simulate the $dI/dV$ spectrum of a single molecular junction in an STM setup[192]. The junction is described by the two-impurity Anderson model with a small magnetic anisotropy energy (MAE). The numerical results demonstrate that the subsystem HEOM substantially reduces the number of ADO and significantly saves the computation cost. Moreover, for many-body OQSs involving spin excitations, the $dI/dV$ spectra simulated by the subsystem HEOM with a small $\widetilde{N}$ show a highly agreement with those simulated by the full-space HEOM. This is probably because high-energy eigenstates of $\hat{H}_S$ have little contribution to the actual low-energy spin excitation processes whose energy scale is orders of magnitude lower than those associated with the charge excitations and fluctuations.

## 2.5 Solving Stationary States and Dissipative Dynamics

HEOM-QUICK2 integrates an efficient and accurate quantum propagator to simulate the dissipative dynamics of generic many-body OQSs. Before introducing this newly proposed propagator, we revisit the HEOM presented in Equation (5) and rewrite it in a compact form of[207,208]

$$\dot{\boldsymbol{\rho}} = -i\boldsymbol{\mathcal{L}}\boldsymbol{\rho}. \tag{10}$$

Here, $\boldsymbol{\rho} \equiv \left\{ \rho_{j_1 \cdots j_n}^{(n)}; n = 0, 1 \ldots, L \right\}$ is a state vector that represents RDO and all ADO, and $\boldsymbol{\mathcal{L}}$ is the superoperator which governs the time evolution of $\boldsymbol{\rho}$. Therefore, simulating the dissipative dynamics of many-body OQSs is equivalent to solving the initial-value problem of differential equations of Equation (10), which is realized via the following three-step protocol:

1) Prepare the initial state for the dynamic simulation. HEOM-QUICK2 provides various choices of initial states. Alternatively, one can employ the stationary solver (will be introduced later) to many-body OQSs under the stationary condition and takes the resulting RDO and ADOs as the initial state;

2) Referring to the physical processes to be explored, impose external perturbations, e.g., time-dependent external fields, on the system-environment composite at time $t_0$;

3) Propagate HEOM by the numerical algorithms, such as the 4th-order Runge-Kutta (RK4)[209] or the Chebyshev algorithm[210,211], until reaching a new stationary state or the final time $t_f$. During the simulation, HEOM-QUICK2 generates auxiliary data files that record the RDO and ADO at pre-designated intermediate time instants.

The stationary state is a critical condition in the study of equilibrium and nonequilibrium many-body OQS. One way to solve the stationary state of Equation (5) is to propagates a many-body OQS from a given initial state to the asymptotic limit of $t \to \infty$. However, this strategy is rather expensive, particularly for the Kondo-correlated system because the formation time of Kondo state could be extremely long[212].

When the many-body OQS reaches a stationary state, the time derivative of the state vector is zero, i.e., $\dot{\boldsymbol{\rho}} = 0$, and Equation (10) reduces to a set of coupled linear equations

$$\mathcal{L}\boldsymbol{\rho} = \boldsymbol{b} \tag{11}$$

where $\boldsymbol{b} = \{1,0,\dots,0\}$ is a vector with its first element satisfying the normalization condition of $\text{tr}_s[\rho_s] = 1$. Consequently, a more efficient approach is to directly solve Equation (11) by the optimizer which uses a certain optimization algorithm, such as the biconjugate gradient (BiCG) approach[209], the transpose-free quasiminimal residue (TFQMR) approach[213] or the Jacobi iteration approach[214], to minimize the Euclidean norm $||\mathcal{L}\boldsymbol{\rho} = \boldsymbol{b}||$. The obtained $\boldsymbol{\rho}$ then serves as the initial state for the subsequent time evolution simulations or iterative stationary state calculations. In most cases, these two approaches yield the same stationary results. However, for some systems with bound states (localized states decoupled with the environment), there may exist multiple stationary solutions. In such cases, the stationary solution calculated by the time evolution method depends on the choice of the initial state. Therefore, the initial state should be carefully chosen in practice so that the initial state is reasonably close to the target stationary state in the phase space.

The HEOM method is often applied to study the evolution of stationary-state properties subject to the tuning of energetic parameter of the system or environment. If we have had knowledge of the stationary-state solution, $\boldsymbol{\rho}$, of a known many-body OQS, the problem becomes how to efficiently obtain a new stationary state $\boldsymbol{\rho}'$ for the same system but subject to a small variation. If the new state $\boldsymbol{\rho}'$ satisfies the following linear equations

$$\boldsymbol{\mathcal{L}}'\boldsymbol{\rho}' = \boldsymbol{b} \tag{12}$$

with the new superoperator $\boldsymbol{\mathcal{L}}'$, one can solve Equation (12) by taking $\boldsymbol{\rho}$ as the initial guess of the stationary optimizer.

An alternative approach is based on the perturbative theory and regards the minor change as a perturbation to the known stationary-state solution[192]. The new stationary state $\boldsymbol{\rho}'$ can be expanded as

$$\boldsymbol{\rho}' \approx \sum_{m=0}^{k} \boldsymbol{\rho}_{[m]}. \tag{13}$$

Here, $\boldsymbol{\rho}_{[0]} = \boldsymbol{\rho}$ is the zeroth-order approximation and $\boldsymbol{\rho}_{[m]}$ is the $m$th-order response, which satisfies the following equation

$$\boldsymbol{\mathcal{L}}'\boldsymbol{\rho}_{[m]} + \Delta\boldsymbol{\mathcal{L}}\,\boldsymbol{\rho}_{[m-1]} = 0 \tag{14}$$

where $\Delta\boldsymbol{\mathcal{L}} \equiv \boldsymbol{\mathcal{L}}' - \boldsymbol{\mathcal{L}}$ is the difference between the hierarchical superoperators with and without the perturbation. If the sum of all $\boldsymbol{\rho}_{[m>k]}$ is negligibly small, the series can be truncated at a sufficiently high order $k$ and the new stationary state is obtained approximately by iteratively solving the linear problem of Equation (14) until the convergence criterion is satisfied.

## 2.6 Acquiring System Observables and Response Properties

With the knowledge of system states, we can evaluate a variety of local observables and response properties for many-body OQSs. The expectation value of any operator $\hat{O}(t)$ with arbitrary time dependence is calculated by tracing over the degrees of freedom of the system-environment composite

$$\langle \hat{O}(t) \rangle = \text{tr}_\text{T}[\hat{O}\rho_\text{T}(t)], \tag{15}$$

where $\rho_\text{T}(t)$ denotes the time-dependent density operator of the total system. For local system operators, Equation (15) reduces to $\langle \hat{O}(t) \rangle = \text{tr}_S[\hat{O}\rho_S(t)]$; Meanwhile, ADOs are involved in calculating the expectation values of operators with reservoir degrees of freedom. For example, the spin-polarized electric current flowing into reservoir-$\alpha$ is evaluated as[80]

$$I_{\alpha s} \equiv \frac{d\langle \hat{N}_{\alpha s}(t) \rangle}{dt} = -2 \sum_{\nu,p} \text{Im}\{\text{tr}_S[\hat{a}_{\nu s}\rho^+_{\alpha\nu ps}]\}, \tag{16}$$

where $\hat{N}_{\alpha s} = \sum_k \hat{d}^\dagger_{\alpha k s}\hat{d}_{\alpha k s}$ and $\{\rho^+_{\alpha\nu ps}\}$ are the first-tier ADO.

HEOM-QUICK2 is also capable of evaluating system response properties by using the linear response theory within the framework of HEOM[215]. A representative example is the system correlation function

$$C_{\hat{A}\hat{B}}(t) \equiv \langle \hat{A}(t)\hat{B}(0) \rangle = \langle\!\langle A(0) | \mathcal{G}(t) | B\boldsymbol{\rho}^\text{st} \rangle\!\rangle. \tag{17}$$

Since the linear response properties of the systems involve inner products in HEOM space, for clarity we introduce the 'superket' and 'superbra' notations used in Liouville space. $|\boldsymbol{\rho}^\text{st}\rangle\!\rangle \equiv \{\rho^{(n)}_{j_1\ldots j_n}; n = 0,1\ldots,L\}$ is a superket which represents the stationary state of many-boy OQSs. $\langle\!\langle A| \equiv \{\hat{A}, 0, \ldots\}$ and $|B\boldsymbol{\rho}^\text{st}\rangle\!\rangle \equiv \{\hat{B}\rho^{(0)}, \hat{B}\rho^{(1)}_{j_1}, \ldots\}$ are superbra and superket with $\hat{A}$ and $\hat{B}$ being two system operators, respectively. $\mathcal{G}(t) = e^{-i\mathcal{L}t}$ is the propagator for state vectors. The frequency-resolved system correlation function is obtained by a half-Fourier transform, i.e., $\tilde{C}_{\hat{A}\hat{B}}(\omega) = \int_0^\infty C_{\hat{A}\hat{B}}(t) e^{i\omega t} dt$.

In practice, HEOM-QUICK2 provides two protocols to evaluate $\tilde{C}_{\hat{A}\hat{B}}(\omega)$: The time-domain protocol is based on the evolution of $C_{\hat{A}\hat{B}}(\omega)$, and proceeds as follows:

1) Employ the stationary solver to obtain the stationary state $|\boldsymbol{\rho}^\text{st}\rangle\!\rangle$ of many-body OQSs;
2) Take the superket $|B\boldsymbol{\rho}^\text{st}\rangle\!\rangle$ as the initial state for the time evolution with the propagator $\mathcal{G}(t)$;
3) Calculate the inner product between superkets and superbras on the right-hand side of Equation (17) and perform the half-Fourier transform.

Alternatively, the frequency-domain protocol utilizes the following relation

$$\tilde{C}_{\hat{A}\hat{B}}(\omega) = \int_0^\infty \langle\langle A(0)|\mathcal{G}(t)e^{i\omega t}|B\rho^{st}\rangle\rangle dt$$

$$= i\langle\langle A(0)|(\omega - \mathcal{L})^{-1}|B\rho^{st}\rangle\rangle$$

$$= \langle\langle A(0)|X(\omega)\rangle\rangle. \tag{18}$$

For a given frequency $\omega$, the superket $|X(\omega)\rangle\rangle$ is the solution to the following linear equations

$$(\omega - \mathcal{L})|X(\omega)\rangle\rangle = i|B\rho^{st}\rangle\rangle. \tag{19}$$

In HEOM-QUICK2, $\tilde{C}_{\hat{A}\hat{B}}(\omega)$ is obtained by using the optimization algorithms to solve Equation (19) and then calculating the inner product in Equation (18). It is noted that these two protocols are equivalent in the linear response regime.

To study the time-dependent response properties $\tilde{C}_{\hat{A}\hat{B}}(\omega, t) \equiv \int_0^\infty \langle \hat{A}(t+\tau)\hat{B}(\tau)\rangle e^{i\omega\tau} d\tau$, HEOM-QUICK2 provides the following protocol: At a given time $t$, replace the stationary state vector $|\rho^{st}\rangle\rangle$ in Equations (18) and (19) by time-dependent $|\rho(t)\rangle\rangle$ obtained from the dynamic simulator, then solve the new linear equations Equation (19) for different $\omega$ and finally obtain $\tilde{C}_{\hat{A}\hat{B}}(\omega, t)$.

## 3. CODE ARCHITECTURE

### 3.1 Overview of HEOM-QUICK2 and its Program Framework

HEOM-QUICK2 is a Fortran-based open-source program for simulating the dynamics of open quantum systems and is freely available for use and/or modification on the Linux platform. Most codes follow the Fortran95 standard, except the codes of the TFQMR iteration algorithm in the F77 format. The library of HEOM-QUICK2 depends on BLAS (Basic Linear Algebra Subprograms) and LAPACK (Linear Algebra PACKage). The program also supports multi-threading with the OpenMP framework to boost matrix calculations on shared memory computers. Therefore, for the compilation of the parallel version of HEOM-QUICK2, the following softwares or libraries are mandatory: Fortran compiler; an implementation of OpenMP; numerical libraries including BLAS and LAPACK. We recommend the latest Intel oneAPI Base & HPC Toolkit or Intel Parallel Studio XE Cluster. These toolkits include the necessary Intel Math Kernel Library and Intel Fortran Compiler with OpenMP support to build the parallel program. HEOM-QUICK2 can be compiled after setting up the options of makefile, e.g. modifying the install location and links to libraries, and finally creates an executable file.

HEOM-QUICK2 follows a procedural programming paradigm, decomposing the workflow of numerically solving HEOM into three modules: the input-output module, preparation module, and calculation module. Each module consists of various subroutines, which are sequentially called in the main program. Figure 6 shows the standard workflow for solving HEOM of generic many-body OQSs. HEOM-QUICK2 first reads input information (e.g. system and environmental parameters, external fields, job control tags, etc.) through the IO module, and feeds them to the preparation module where these input parameters are processed and necessary information about the setting of simulation is generated for subsequent computations. The computation module implements accurate and efficient algorithms to solve stationary states and dissipative dynamics for the given system. The program also employs the obtained RDO and ADOs to calculate local observables and response properties. Finally, the IO module generates auxiliary files which save the details of the above workflow and calculated results.

**3.2 Input/Output Module**

Based on the Fortran95 coding standards, the I/O module provides a specified I/O format that achieves effective communication between users and the program. The I/O module begins with reading the parameters and tags from the input file that user carefully prepared. Throughout the execution of the program, the I/O module records the processing details of the preparation and calculation modules and the generated intermediate results. This allows users to closely monitor the progress and performance of the program. Finally, the I/O module presents the computed HEOM vector $\rho$ and various system properties, providing valuable outcomes to users.

The input file is a tagged format ASCII file. Figure 7(a) and (b) show the input files for preparing an equilibrium state of a SIAM and the sequent simulating the real-time electric current driven by an *ac* voltage, respectively. These input files include the parameters of system Hamiltonian, statistical properties of reservoirs, the information of external fields and job control.

The system Hamiltonian $\hat{H}_S(t)$ depends on the target many-body OQS to be studied. For the case depicted in Figure 7, the parameters of system Hamiltonian are defined by the Fortran namelist "para1" which includes impurity levels and Columb interaction of SIAM. HEOM-QUICK2 has encoded a number of namelists which describe different types of system Hamiltonian. Necessary statistical properties of reservoirs are defined in Lines 7-10 of the input files, including the band width $W_\alpha$, the system-reservoir hybridization $\Gamma_\alpha$, the reservoir temperature $T_\alpha$ and the spin-specific chemical potential $\mu_{s\alpha}$.

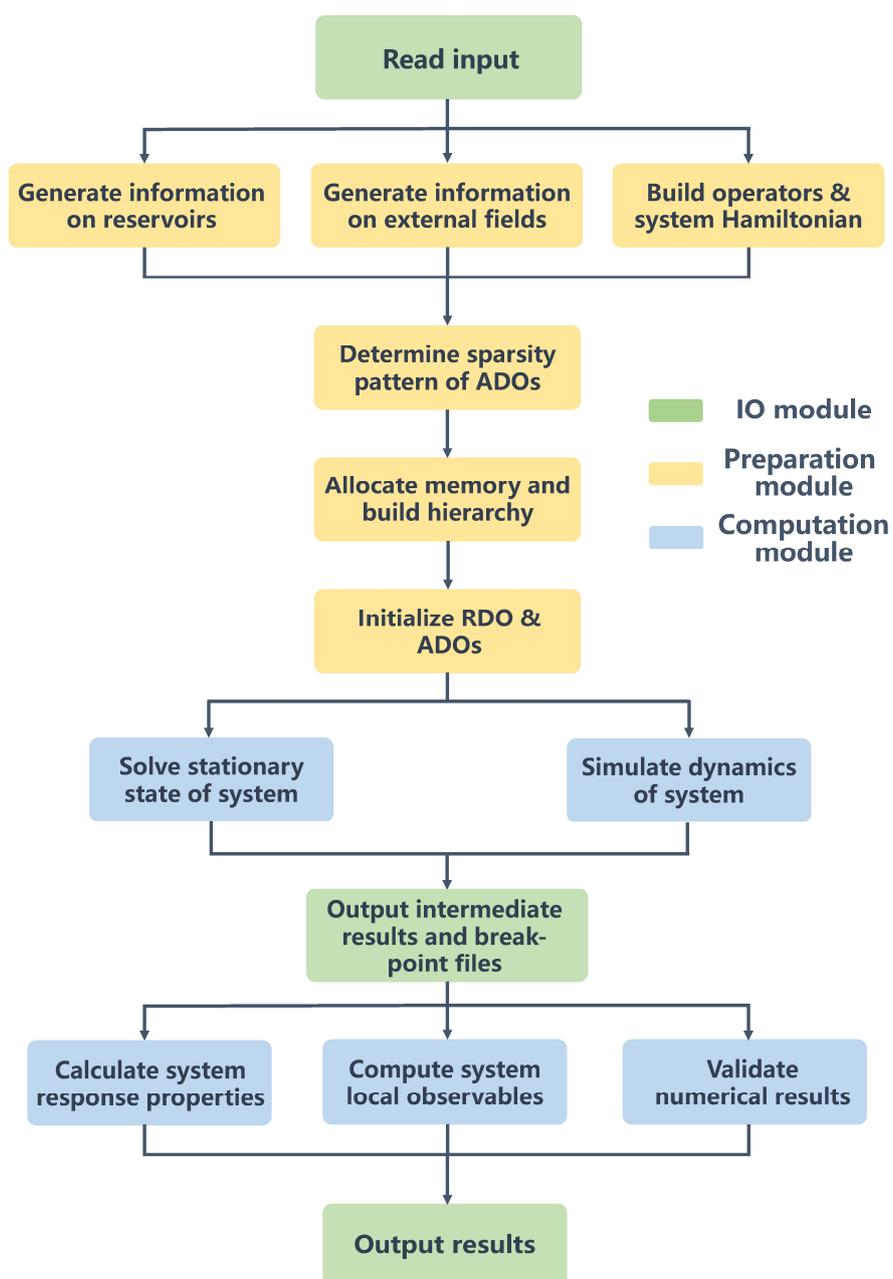

Figure 6: The workflow of HEOM-QUICK2. The program consists of the IO (green), preparation (yellow) and computation (blue) modules. First, the program reads input information through the IO module, and feeds them to the preparation module where these input parameters are processed and necessary information about the system, environment and external fields is generated for subsequent computations. The computation module implements accurate and efficient algorithms to solve the stationary state and/or dissipative dynamics. The program also evaluates local observables and response properties of the system and validates numerical results. Finally, the IO module records the details of the above workflow and outputs intermediate and final results in auxiliary data files.

The namelist "field" offers various types of time-dependent bias voltage. For the stationary calculation in Figure 7(a), the constant bias voltage is simulated by an exponential voltage $V_{s\alpha}(t) = V_{s\alpha}[1 - \exp(-t/\tau_{s\alpha})]$ with a sufficiently small characteristic time $\tau_{s\alpha}$, and the sinusoidal voltage $V_{\alpha s}(t) = V_{\alpha s}\sin(\omega_{\alpha s} t)$ with the frequency $\omega_{\alpha s}$ is employed to describe the *ac* voltage in Figure 7(b).

Job control is the center of the input file, which determines the details of HEOM calculations and has a significant impact on the accuracy of numerical outcomes, and therefore needs to be controlled by multiple tags and namelists together. The tag in Line 1 represents the calculation type with "1" denoting dynamic simulation and "2" being stationary state calculation, respectively. Lines 2 and 3 determine the truncation tier $L$ and the number of basis functions in the Matsubara or Padé scheme. The following Lines 4-6 offer the number of impurities $N_v$, the spin degrees of freedom $N_s$ and the number of reservoirs $N_\alpha$. Moreover, "jobinfo" specifies the spectrum decomposition scheme and the utilization of the sparse matrix technology. The utilization of the adiabatic terminator is declared by the "lad" tag in "adiabatic".

The "method" namelist defines the numerical algorithm for iterative solving the stationary state of a many-body OQS and its convergence criterion and the maximum iteration step are given in "converge". TFQMR is the default solver defined by the tag "methodss=2". Other iterative methods are also available in the program, such as the BiCG ("methodss=0") and Jacobi methods ("methodss=4"). The time evolution algorithm is specified in "tdjob" and includes the 4th-order Runge-Kutta method ("tdmethod=0") and the Chebyshev polynomial expansion method ("tdmethod=1"). These details and implementation of these algorithms are presented later in Sec. 3.4.

When studying electronic structures of realistic many-body OQSs, one can extract the energetic parameters of the system Hamiltonian and reservoirs from quantum chemistry calculations. For example, the system-reservoir coupling can be extracted from the DFT calculation on projected density of states[141,142,162], and, for example, MAE and the spin-spin exchange interaction of a nanomagnet can be calculated by the post Hartree-Fock methods[140,143,161,216].

```
(a)                                                          (b)
1    2                          ! solve ρˢᵗ                  1    1                          ! simulate ρ(t)
2    4                          ! truncation tier L          2    4                          ! truncation tier L
3    2                          ! number of Padé poles       3    2                          ! number of Padé poles
4    1                          ! DoF of impurity Nᵥ         4    1                          ! DoF of impurity Nᵥ
5    2                          ! DoF of spin Nₛ             5    2                          ! DoF of spin Nₛ
6    2                          ! number of baths Nα         6    2                          ! number of baths Nα
7    5.0d0  5.0d0               ! band width W_L,W_R         7    5.0d0  5.0d0               ! band width W_L,W_R
8    1.0d-1  1.0d-1             ! system-bath coupling Γ_L,Γ_R  8  1.0d-1  1.0d-1           ! system-bath coupling Γ_L,Γ_R
9    5.0d-3  5.0d-3             ! bath temperature T_L,T_R   9    5.0d-3  5.0d-3             ! bath temperature T_L,T_R
10   0.0d0 0.0d0 0.0d0 0.0d0    ! chemical potential μ↑L,μ↓L,μ↑R,μ↓R  10  0.15d0 0.15d0 -0.15d0 -0.15d0  ! μ↑L,μ↓L,μ↑R,μ↓R
11   5.0d2                      ! end time of dynamics t_end 11   5.0d2                      ! end time of dynamics t_end
12   1.0d-2                     ! time step Δt               12   1.0d-2                     ! time step Δt
13   ! energetic parameters of single impurity Anderson model  13  ! energetic parameters of single impurity Anderson model
14   $para1 eup = -0.6d0                                     14   $para1 eup = -0.6d0
15          edown = -0.6d0                                   15          edown = -0.6d0
16          uu = 1.2d0 $end                                  16          uu = 1.2d0 $end
17   ! time-dependent bias voltage μₛₐ(t) = μₛₐ[1 - e^(-t/τₛₐ)]  17  ! time-dependent bias voltage μₛₐ(t) = μₛₐ sin(ωₛₐ t)
18   $field fieldtype = 0 $end                               18   $field fieldtype = 1 lreadomega = .true. $end
19   1.0d-20 1.0d-20 1.0d-20 1.0d-20   ! τ↑L,τ↓L,τ↑R,τ↓R     19   0.06d0 0.06d0 0.06d0 0.06d0   ! ω↑L,ω↓L,ω↑R,ω↓R
20   ! job information                                        20   ! job information
21   ! use sparse matrix technique and Prony fitting scheme  21   ! use sparse matrix technique and Prony fitting scheme
22   ! itype_psf = 5 for Tₑ = 0.005 eV                       22   ! itype_psf = 5 for Tₑ = 0.005 eV
23   $jobinfo lsparse = .true.                               23   $jobinfo lsparse = .true.
24            psfjob = .true. itype_psf = 5 $end             24            psfjob = .true. itype_psf = 5 $end
25   ! convergence criterion for calculations                25   ! convergence criterion for calculations
26   $converge maxit0 = 20000 crit = 1.0d-8 $end             26   $converge maxit0 = 20000 crit = 1.0d-8 $end
27   ! adopt the adiabatic terminator                        27   ! adopt the adiabatic terminator
28   $adiabatic lad = .true. $end                            28   $adiabatic lad = .true. $end
29                                                            29   ! output tapefiles after nresume = 100 steps
30   ! use TFQMR as iterative solver                         30   ! only active for time evolution
31   $method methodss = 2 $end                               31   $resume icount = 0 lresume= .true. nresume = 5000 $end
32                                                            32   ! use 4th-order Runge-Kutta for time propagation
33                                                            33   $tdjob tdmethod = 0 $end
```

Figure 7: The standard input files for (a) solving an equilibrium state of SIAM and the subsequent dissipation dynamics driven by a constant bias voltage.

The I/O module generates a variety of output files in ASCII or binary format, which serve to record the calculation process on-the-fly, store intermediate outcomes during the computation, and preserve the final results. The main output file is in ASCII format and records the details of the entire workflow throughout the running of the HEOM-QUICK2 program. This includes a copy of the input file, information on the spectrum decomposition of reservoir correlation functions, the system Hamiltonian and its eigenvalues, the initial RDO and ADOs, details of the iterative process for solving the stationary state or time steps for dissipative dynamics simulations, and other important system properties.

Necessary auxiliary output files are created within the subsequent preparation and calculation modules. The preparation module generates binary files for storing hierarchy information. For example, the output index table is used to search for a given ADO in the binary 'indextable' file, and the coefficients $\{\eta_{j\nu}\}$ as well as the dissipation rates $\{\gamma_j\}$ of principal dissipation modes involved in the upper- and lower-tier operations of each ADO are stored in the binary 'coefindex' file. Furthermore, a Python program is offered to read the coefficients $\{\eta_{j\nu}\}$ and dissipation rates $\{\gamma_j\}$ given by the decomposition scheme saved in a specific ASCII file. The computational module automatically generates a sequence of breakpoint files in binary format to store intermediate results during the iterative solution of stationary states or in the simulation of dissipative dynamics using time propagation methods. These files serve as essential references for the accurate execution and analysis of the calculations, and the Python programs are offered for necessary post-processing. This ensures comprehensive documentation of the system dynamics, enabling detailed analysis and examination of the simulation results. At the end of calculations, the resulting HEOM vector $\boldsymbol{\rho}$ is stored in a dedicated binary file, which can serve as an initial state for subsequent or continued jobs.

**3.3 Preparation Module**

The preparation module processes the information passing from the IO module. In the preparation module, HEOM-QUICK2 first employs the spectrum decomposition scheme defined in the input file to unravel reservoir correlation functions $C(t)$ into a series of exponential functions, and then compares the discrepancy between the decomposition results and the exact values. The Prony fitting scheme has been already performed at different temperatures and the resulting parameters are stored in the HEOM-QUICK2 presets for users. The Prony scheme employed here adopts a cut-off time much longer than the inverse of the temperature, i.e. $t_{\text{cut}} \gg 1/T$, which exceeds the ending time of most dissipative dynamics. Moreover, HEOM-QUICK2 also provides the Fortran and Python codes of the Prony scheme for any temperature, enabling users to unravel $C(t)$ as needed.

The frequency-domain approaches, such as the Matsubara, Padé and Fano schemes will be performed immediately at the start of HEOM calculations. Particularly for the Fano scheme, the program provides three sets of fitting coefficients of the generalized Fano functions. All these sets can be adopted directly for any temperature $T$ without refitting.

If the system parameters change in continuation calculations while the environmental temperatures remain constant, the program will not implement the Prony scheme again but will directly adopt the preset parameters. In contrast, the program will quickly recalculate the decomposition parameters given by the frequency-domain schemes, even if the temperature does not change.

Based on the specific number of impurities $N_v$ and spin degrees of freedom $N_s$, the program constructs the system creation (annihilation) operators and spin operators in matrix form with the correct dimensions. The system Hamiltonian can be thus explicitly built by the matrix product of these operators and the given energetic parameters.

The program provides a variety of external fields to study the effect of external perturbations on the dissipation dynamics or stationary state of many-body OQSs. For example, the influence of temperature gradient or bias voltage is manifested by the modified reservoir correlation functions, which effectively changes the parameters obtained from the spectrum decomposition. Local external magnetic fields and time-dependent gate voltages imposed on local systems lead to additional terms in the system Hamiltonian.

HEOM-QUICK2 performs the vectorization of all system operators and ADOs, and maps HEOM in Equation (1) from a matrix-matrix equation into a matrix-vector equation in Equation (10). The coordinate (COO)-format sparse matrix technique is employed to significantly reduce the requirement of physical memory to only 5%~10% of those needed in the dense matrix format[197]. The detailed benchmarks can be found in the previous review[157]. To determine the dimension of the vectorized HEOM state $\rho$, HEOM-QUICK2 first evaluates the sparsity pattern of ADOs at different truncation tiers, which can be determined by the sparsity pattern of $\hat{H}_S$ and the hierarchy structure together[157]. The program enumerates all possible combinations of principal dissipation modes in ADOs and the hierarchy is built by indexing the up-tier and down-tier operators on RDO and ADOs in each tier. The fermionic anti-commuting relation is fully taken into account to fulfil the correct fermionic statistics. The program counts the total number of nonzero elements in RDO and ADOs, and allocates the physical memory required for storage. It is noted that HEOM-QUICK2 does not explicitly construct the HEOM superoperator $\mathcal{L}$ in Equation (10) to update the HEOM state $\rho$, but instead calculates RDO and each ADO based on Equation (1). Finally, RDO and ADOs are initialized according to the job control tag before entering into the computation module.

The preparation modules only support single-thread processing. Therefore, for many-body OQSs with multiple degrees of freedom, it is recommended to carefully choose the appropriate truncation tier and the number of basis functions so as to avoid a too large hierarchy structure that significantly increase the computation time.

## 3.4 Computation Module

Following the task type and numerical algorithm specified in the input file, the computation module takes RDOs and ADOs passed by the preparation module as the initial state and solves the stationary state or dissipation dynamics of many-body OQSs. The calculated HEOM state $\rho$ will be used to calculate response properties and local observables.

HEOM time evolution is equivalent to numerically solving the initial value problem for ordinary differential equations (ODEs) of the vectorized HEOM in Equation (10)

$$\dot{\rho} = -i\mathcal{L}\rho$$

with a known initial state $\rho(t_0)$. Since HEOM-QUICK updates the state $\rho$ by calculating RDO and each ADO based on Equation (5), the matrix operations for different ADOs can be parallelized on multi-CPUs using OpenMP, which substantially improves the efficiency of HEOM computation.

The program provides the 4th-order Runge-Kutta (RK4) method[209] and the Chebyshev polynomial expansion method[210,211] for time propagation. The Runge-Kutta method is a single-step method motivated by the Taylor expansion technique. This method does not require derivatives of the right-hand side in ODEs and therefore serves as general-purpose initial value problem solver. RK4, due to its numerical robustness and reasonably simple implementation, has become an effective and widely used method for solving the initial value problems of ODEs. The time interval $[t_0, t_f]$ is divided into $N_t$ evenly spaced time instants $\{t_i; i = 1, ..., N_t\}$ with a time interval $dt \equiv (t_f - t_0)/N_t$. The RK4 method is defined by using the following recursion formula:

(1) Start:

    (a) Set the final time $t_f$ and the time interval $dt$;

    (b) Prepare the initial state $\rho(t_0)$.

(2) For $i = 1, ..., N_t$ do:

    (a) Compute $\boldsymbol{k}_1 = \mathcal{L}\,\boldsymbol{\rho}(t_i)$, $\boldsymbol{k}_2 = dt * \mathcal{L}\,(\boldsymbol{\rho}(t_i) + 0.5 * \boldsymbol{k}_1)$, $\boldsymbol{k}_3 = dt * \mathcal{L}\,(\boldsymbol{\rho}(t_i) + 0.5 * \boldsymbol{k}_2)$, and $\boldsymbol{k}_4 = dt * \mathcal{L}\,(\boldsymbol{\rho}(t_i) + \boldsymbol{k}_3)$;

    (b) Update $\boldsymbol{\rho}(t_{i+1}) = \boldsymbol{\rho}(t_i) + (\boldsymbol{k}_1 + \boldsymbol{k}_2 + \boldsymbol{k}_3 + \boldsymbol{k}_4)/6$.

An alternative numerical method, the Chebyshev polynomial expansion, is a global propagator method based on the fact that HEOM in Equation (10) is a first order homogeneous linear ODE and has the formal solution of

$$\boldsymbol{\rho}(t) = \exp(-i\mathcal{L}t)\,\boldsymbol{\rho}(t_0),$$

when the superoperator $\mathcal{L}$ is time-independent. The main idea is to use a Chebyshev polynomial expansion of the evolution operator

$$\exp(-i\mathcal{L}t) = \sum_{n=0}^{N_b} a_n(t)\Phi(-i\mathcal{L}). \tag{20}$$

Here, $\{a_n(t)\}$ represents a set of time-dependent polynomial coefficients and $\Phi(\mathcal{L})$ is the complex Chebyshev polynomials as a function of the operator. Due to the range of definition of Chebyshev polynomials, the superoperator has to be normalized by dividing by $\Delta L_{\text{grid}} = L_{\max} - L_{\min}$, and can be shifted to

$$\mathcal{L}_{\text{norm}} = \frac{2\mathcal{L} - (L_{\max} + L_{\min})}{\Delta L_{\text{grid}}}. \tag{21}$$

Here, $L_{\max}$ and $L_{\min}$ in principle represent the maximum and minimum eigen-values of $\mathcal{L}$, and are approximated by the maximum and minimum elements of $-i\mathcal{L}_s - \sum_j \gamma_j$ of all ADOs if $\mathcal{L}$ is diagonal-dominated. The time-dependent HEOM vector can be expanded as[210,211]

$$\boldsymbol{\rho}(t) \approx e^{-\frac{i(L_{\max}+L_{\min})t}{2}} \sum_{n=0}^{N_b} a_n(\alpha)\boldsymbol{\phi}_n.$$

Here, $\boldsymbol{\phi}_n = \Phi_n(-i\mathcal{L}_{\text{norm}})\boldsymbol{\rho}(t_0)$ are defined by the recursion relation of Chebyshev polynomials, $\boldsymbol{\phi}_{n+1} = -2i\mathcal{L}_{\text{norm}}\boldsymbol{\phi}_n + \boldsymbol{\phi}_{n-1}$. The recurrence begins with $\boldsymbol{\phi}_0 = \boldsymbol{\rho}(t_0)$ and $\boldsymbol{\phi}_1 = -i\mathcal{L}_{\text{norm}}\boldsymbol{\rho}(t_0)$. The expansion coefficients $\alpha_n$ can be calculated by the Bessel function of the first kind, $\alpha_n(\alpha) = 2J_n(\alpha)$ with $\alpha_0(\alpha) = J_0(\alpha)$ and $\alpha = \Delta E_{\text{grid}}\, t/2$. It is found that the Bessel function $J_n$ decays exponentially when $n > \alpha$. Therefore, the number of expansion terms can be estimated by specifying the criterion precision of $J_n$. The Chebyshev polynomial expansion method proceeds as follows[210,211]:

(1) Start:
  (a) Set the final time $t_f$ and the time interval $dt$;
  (b) Prepare the initial state $\boldsymbol{\rho}(t_0)$;
  (c) Define the expansion precision of $J_n$ and estimate the number of expansion term $N_b$;
  (d) Find $L_{\max}$ and $L_{\min}$, and then evaluate the factor $\Delta L_{\text{grid}} = L_{\max} - L_{\min}$;
  (e) Renormalize the HEOM superoperator $\mathcal{L}_{\text{norm}} = \frac{2\mathcal{L}-(L_{\max}+L_{\min})}{\Delta L_{\text{grid}}}$.

(2) For $i = 1, \dots, N_t$ do:

(a) Calculate $\boldsymbol{\phi}_1 = -i\mathcal{L}_{\text{norm}}\boldsymbol{\rho}(t_0)$;

(b) For $n = 1, \ldots, N_b$ do:

   a) Update $\boldsymbol{\phi}_{n+1} = -2i\mathcal{L}_{\text{norm}}\boldsymbol{\phi}_n + \boldsymbol{\phi}_{n-1}$;

   b) Determine the expansion coefficient $a_n(\alpha)$ with $\alpha = \Delta E_{\text{grid}}\, t/2$;

(c) Calculate $\boldsymbol{\rho}(t) \approx e^{-\frac{i(L_{\max}+L_{\min})t}{2}} \sum_{n=0}^{N_b} a_n(\alpha)\,\boldsymbol{\phi}_n$.

We apply the RK4 and the Chebyshev methods to study the dissipation dynamics of SIAM driven by a constant bias voltage. It is noted that the Chebyshev method has not yet been combined with any sparse matrix technique. Therefore, instead of the sparse matrix technique, the dense matrix technique is adopted in the RK4 calculation. Initially, the impurity system is prepared in an equilibrium state whose population obeys the Boltzmann distribution. A gate voltage is imposed on the system after $t \geq t_0$, which effectively shifts the impurity level and induces the dissipation dynamics. Figure 8 shows the time evolution of $s$-spin electron occupation numbers $n_{s=\uparrow,\downarrow}$. Apparently, the resulting curves of both methods agree reasonably well with each other. Furthermore, the RK4 method yields a more efficient HEOM computation, as manifested by the physical memory costed by two methods in Figure 8. Due to the superior efficiency, the RK4 method is chosen as the default dynamic solver for HEOM-QUICK2 and supports the COO sparse matrix technique and the adiabatic terminator in Sec. 2. Another important function of the calculation module is to solve the stationary state problem of HEOM. One feasible option is to real-time propagate HEOM of Equation (10) by using the above numerical algorithms until the stationary state is reached. The main drawback of this strategy is that the relaxation of the system to a stationary state tends to require a long propagation time, e.g. in the case of strong non-Markovian effect.

An alternative strategy is to directly solve the steady-state HEOM of Equation (11)

$$\mathcal{L}\boldsymbol{\rho} = \boldsymbol{b}.$$

This corresponds to solving a set of linear equations. HEOM-QUICK2 provides different iterative techniques such as the biconjugate gradient (BiCG)[209], the transpose-free quasi-minimal residual (TFQMR) approach[213] and the blocked Jacobi method[214], which trading off accuracy and runtime for solving large equations.

The BICG method allows for solving linear, but not necessarily positive-definite or symmetric, equations. The procedure of the BiCG method can be stated follows[209]:

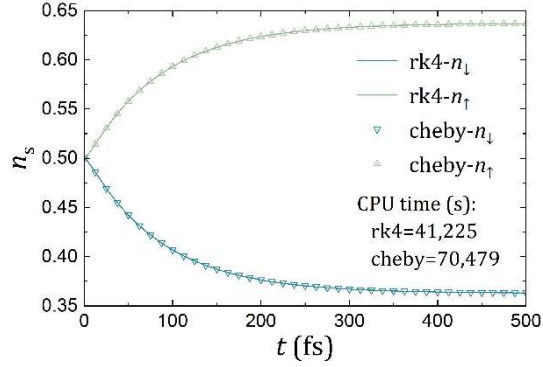

Figure 8: The time evolution of occupation numbers $n_{s=\uparrow,\downarrow}$ of SIAM subjected to a constant bias voltage by using the 4th-order Runge-Kutta and Chebyshev polynomial expansion methods, respectively. Here, the Chebyshev method uses 50 Bessel functions to reach a precision less than 1E-21. HEOM-QUICK2 employs the zero-value scheme to truncate the hierarchy at $L = 3$ and adopts the Prony decomposition scheme with 9 poles. The benchmark was done on a workstation with Intel(R) Xeon(R) Gold 6248R CPU @ 3.00GHz with 64 GB memory. 24 CPUs were used for the parallel solution to HEOM equations and the CPU times cost by each method are list. Both methods require the same amount of memory. The energetic parameters adopted here are (in the unit of eV): $\epsilon_\uparrow = \epsilon_\downarrow = -U/2 = -1.0, T = 0.04, V_\uparrow = -0.01, V_\downarrow = 0.01, \Gamma = 0.15$ and $W = 5$.

(1) Start:

    (a) Choose an initial $\boldsymbol{\rho}_0$, the maximum iterative step $N_{\max}$ and the criterion;

    (b) Evaluate the residual $\boldsymbol{r}_0 = \mathcal{L}\boldsymbol{\rho}_0 - \boldsymbol{b}$;

    (c) Initialize the vectors $\boldsymbol{q}_0 = \boldsymbol{s}_0 = \boldsymbol{p}_0 = \boldsymbol{r}_0$ and the parameters $\alpha_0 = \beta_0 = 1$.

(2) For $i = 1, 2, ..., N_{\max}$ do:

    (a) $\alpha_i = \boldsymbol{s}_{i-1}^H \cdot \boldsymbol{r}_{i-1} / \boldsymbol{q}_{i-1}^H \cdot \mathcal{L} \cdot \boldsymbol{p}_{i-1}$;

    (b) $\boldsymbol{\rho}_i = \boldsymbol{\rho}_{i-1} + \alpha_i \boldsymbol{p}_{i-1}$;

    (c) $\boldsymbol{r}_i = \boldsymbol{r}_{i-1} - \alpha_i \mathcal{L} \boldsymbol{p}_{i-1}$;

    (d) $\boldsymbol{s}_i = \boldsymbol{s}_{i-1} - \alpha_i \mathcal{L}^H \boldsymbol{q}_{i-1}$;

    (e) $\beta_i = \boldsymbol{s}_i^H \cdot \boldsymbol{r}_i / \boldsymbol{s}_{i-1}^H \cdot \boldsymbol{r}_{i-1}$;

    (f) $\boldsymbol{p}_i = \boldsymbol{r}_i + \beta_i \boldsymbol{p}_{i-1}$;

    (g) $\boldsymbol{q}_i = \boldsymbol{s}_i + \beta_i \boldsymbol{q}_{i-1}$.

Here, the superscript $H$ denotes the Hermit conjugation. As an early Krylov subspace method, the BiCG method has a simple numerical implementation, but may yield irregular convergence behaviour[217]. Consequently, BiCG is rarely employed in the present HEOM-QUICK running and has not been integrated with sparse matrix techniques and the new methods developed recently.

To enhance the robustness of iterative algorithms, the TFQMR method has been developed and proceeds as follows[213]:

(1) Start:

    (a) Choose an initial $\boldsymbol{\rho}_0$, the maximum iterative step $N_{\max}$ and the criterion;

    (b) Initialize the vectors $\boldsymbol{p}_0 = \boldsymbol{u}_0 = \boldsymbol{r}_0 = \boldsymbol{b} - \mathcal{L}\boldsymbol{\rho}_0$, $\boldsymbol{v}_0 = \mathcal{L}\boldsymbol{p}_0$ and $\boldsymbol{d}_0 = 0$;

    (c) Choose $\bar{\boldsymbol{r}}_0$ such that $\xi_0 = \bar{\boldsymbol{r}}_0^H \cdot \boldsymbol{r}_0 \neq 0$;

    (d) Set $\tau_0 = \omega_0 = \|\boldsymbol{r}_0\|$, $\theta_0 = \eta_0 = 0$.

(2) For $i = 1, 2, \ldots, N_{max}$ do:

    (a) $\sigma_{i-1} = \bar{\boldsymbol{r}}_0^H \cdot \boldsymbol{v}_{i-1}$, $\alpha_{i-1} = \xi_{i-1}/\sigma_{i-1}$;

    (b) $\boldsymbol{q}_i = \boldsymbol{u}_{i-1} - \alpha_{i-1}\boldsymbol{v}_{i-1}$;

    (c) $\boldsymbol{r}_i = \boldsymbol{r}_{i-1} - \alpha_{i-1}\mathcal{L}(\boldsymbol{u}_{i-1} + \boldsymbol{q}_i)$;

    (d) $\gamma_i = \boldsymbol{r}_i^H \cdot \boldsymbol{r}_i$, $\xi_i = \bar{\boldsymbol{r}}_0^H \cdot \boldsymbol{r}_i$, $\beta_i = \xi_i/\xi_{i-1}$;

    (e) For $m = 2i - 1, 2i$ do:

        1) $\theta_m = \omega_{m+1}/\nu_{m-1}$ with $\omega_{m+1} = \sqrt{\gamma_i}$ if $m = 2i$ and $\omega_{m+1} = (\gamma_{i-1}\gamma_i)^{1/4}$ if $m = 2i - 1$;

        2) $c_m = 1/\sqrt{1 + \theta_m^2}$; $\tau_m = \tau_{m-1}\theta_m c_m$; $\eta_m = c_m^2 \alpha_{n-1}$;

        3) $\boldsymbol{d}_m = \boldsymbol{y}_m + (\theta_{m-1}^2 \eta_{m-1}/\alpha_{m-1})\boldsymbol{d}_{m-1}$ with $\boldsymbol{y}_m = \boldsymbol{u}_{i-1}$ if $m = 2i$ and $\boldsymbol{y}_m = \boldsymbol{q}_i$ if $m = 2i - 1$;

        4) $\boldsymbol{\rho}_m = \boldsymbol{\rho}_{m-1} + \eta_m \boldsymbol{d}_m$;

    (f) $\boldsymbol{u}_i = \gamma_i + \beta_i \boldsymbol{q}_i$; $\boldsymbol{p}_i = \boldsymbol{u}_i + \beta_i(\boldsymbol{q}_i + \beta_i \boldsymbol{p}_{i-1})$; $\boldsymbol{v}_j = \mathcal{L}\boldsymbol{p}_j$.

In comparison to BiCG, TFQMR generally has much smoother convergence, although it may be slightly less efficient due to the additional matrix-vector operations required at each step[217]. TFQMR serves as the default iterative solver for the stationary state of HEOM, supporting sparse matrix operations and compatible with the new methods mentioned in Sec. 2.

The Jacobi method is an algorithm for determining the solutions of a diagonally dominant linear equations. To perform the Jacobi method on HEOM, we revisit the steady-state HEOM of Equation (11) written as

$$0 = \left(-i\mathcal{L}_s + \sum_r \gamma_{j_r}\right) \rho^{(n)}_{j_1 \cdots j_n} + \sum_j \mathcal{A}_j \, \rho^{(n+1)}_{j j_1 \cdots j_n} + \sum_r \mathcal{C}_{j_r} \, \rho^{(n-1)}_{j_1 \cdots j_{r-1} j_{r+1} \cdots j_n}.$$

| Method | $n_\uparrow$ | $n_\downarrow$ | Physical memory (MB) | CPU time (s) |
|---|---|---|---|---|
| RK4 | 0.588729 | 0.411271 | 33.6 | 1,301,175 |
| Jacobi | 0.588724 | 0.411276 | 30.8 | 2,970 |
| TFQMR | 0.588729 | 0.411271 | 50.9 | 1,992 |

Table 3: HEOM-QUICK2 employs the RK4 time propagation method, the Jacobi and TFQMR iterative methods to solve a nonequilibrium stationary state of a SIAM subjected to a constant bias voltage. The benchmark was done on a workstation with Intel(R) Xeon(R) Gold 6248R CPU @ 3.00GHz with 64 GB memory. 24 CPUs were used for the parallel solution to HEOM, and the convergence criterion is the same for two iterative methods. The Jacobi method adopts $\epsilon = 2.0$ and the final time of time propagator is $t_f = 2,069$ at which the first six significant digits of RDO obtained by RK4 are equal to the results of TFQMR. Here lists electron occupancy $n_\uparrow$ and $n_\downarrow$, physical memory (in unit of MB), and CPU time (in unit of second) required for calculations. The hierarchy is truncated by the zero-value scheme at $L = 4$ and by the Prony scheme with 12 poles. The energetic parameters adopted here are (in the unit of eV): $\epsilon_\uparrow = \epsilon_\downarrow = -U/2 = -1$, $T = 0.003$, $\Gamma = 0.02$, $V_\uparrow = -V_\downarrow = 0.001$ and $W = 5.0$.

This equation can be further recast as

$$\rho^{(n)}_{j_1 \cdots j_n} = \left(-i\mathcal{L}_s + \sum_r \gamma_{j_r} + \epsilon\right)^{-1} \left[\epsilon \, \rho^{(n)}_{j_1 \cdots j_n} - \sum_j \mathcal{A}_j \, \rho^{(n+1)}_{j j_1 \cdots j_n} - \sum_r \mathcal{C}_{j_r} \, \rho^{(n-1)}_{j_1 \cdots j_{r-1} j_{r+1} \cdots j_n}\right]. \quad (22)$$

by introducing a relaxation parameter $\epsilon > 0$ which guarantees HEOM is diagonally dominant. This is a self-consistent linear equation that can be solved iteratively. For brevity, we rewrite Equation (5) into a matrix-vector equation, i.e. $\boldsymbol{\rho} = \mathcal{L}^* \boldsymbol{\rho}$ with $\mathcal{L}^*$ describing the new hierarchy in this equation.

Increasing $\epsilon$ improves the numerical stability but decreases the convergence speed. To balance accuracy and efficiency, it is appropriate to set $\epsilon$ as the value about the spectrum span of $\mathcal{L}_s$ in a case of weak system-environment coupling. However, in a case of strong coupling, $\epsilon$ should be tuned up to ensure the numerical stability at the cost of convergence speed. The algorithm of the Jacobi method is stated follows:

(1) Start:

   (a) Choose an initial $\boldsymbol{\rho}_0$, the maximum iterative step $N_{\max}$ and the criterion precision;

(b) Specify the relaxation parameter $\epsilon$.

(2) For $i = 1, 2, ..., N_{\max}$ do:

Update $\boldsymbol{\rho}_i = \mathcal{L}^* \boldsymbol{\rho}_{i-1}$ until $\|\boldsymbol{\rho}_i - \boldsymbol{\rho}_{i-1}\| < crit$.

We carry out a benchmark test to solve a nonequilibrium stationary state of a SIAM subjected to a constant bias voltage by using the RK4 time propagation method, the Jacobi and TFQMR iterative methods, respectively. As shown in Table 3, all three methods yield reasonably accurate electron occupation numbers. The TFQMR solver displays the fastest computation time, followed by the Jacobi method. Although RK4 requires a sufficiently long time evolution to reach a stationary state, it consumes the second least physical memory. TFQMR requires the most additional memory to store necessary auxiliary vectors and matrices. Nevertheless, due to its smooth and quick convergence, we still recommend TFQMR as the first choice to solve the stationary HEOM, provided that the computer has sufficient memory and that TFQMR does not occasionally fail to convergence in calculations.

Recently, Kaspar *et al.* have discussed the performance of different iterative algorithms for solving HEOM[93]. Their numerical experiments indicate that the conjugate gradient on normal equations (CGNE[218]) and least-squares algorithms (LSQR[219] and LSMR[220]) schemes have good convergence properties and remain robust over a wide range of parameters. Based on these tests, they argue that the CGNE algorithm is the best compromise between accuracy, required CPU time, stability, and memory usage. We also focus on the development of iterative algorithms and plan to update more efficient and accurate iterative algorithms in future versions.

Following an analysis of the trace unity of RDO and the positivity of its diagonal elements, the obtained HEOM vector is applied to calculate response properties and local observables of the system. An important issue in HEOM calculations is to find the minimum truncation tier, denoted as $L_{\min}$, at which converged results of the desired system properties can be obtained. In practice, the value of $L_{\min}$ varies depending on the details of the system Hamiltonian and environmental statistical properties. For example, Figure 4(c) and (d) demonstrate that the lowest truncation tiers, $L_{\min}$, required to achieve quantitatively accurate convergence of the zero-frequency impurity spectral functions, vary with the strength of the system-environment coupling. To balance the accuracy and efficiency in HEOM calculations, it is highly recommended for users to perform calculations at different truncation tier $L$ first. This allows them to determine $L_{\min}$ that ensures convergence for the desired system properties.

## 4. FUNCTIONALITY AND NUMERICAL EXAMPLES

### 4.1 Building Quantum Impurity Models

We exemplify the construction of a quantum impurity model for an adsorbed nanomagnet in the SP-STM setup, and the input file for the HEOM-QUICK2 program is shown in Figure 9. The nanomagnet is in the presence of an external magnetic field in the $z$-direction, and a non-zero bias voltage is applied to the tip. The total OQS is described by an extended two-impurity Anderson impurity model, and the system Hamiltonian consists of three parts

$$\widehat{H}_S(t) = \widehat{H}^{\text{ch}} + \widehat{H}^{\text{spin}} + \widehat{H}^{\text{ext}}(t). \tag{23}$$

Here, $\widehat{H}^{\text{ch}}$ determines the system's charge state

$$\widehat{H}^{\text{ch}} = \sum_{\nu,s} \epsilon_{\nu s} \hat{n}_{\nu s} + \sum_{\nu} U_\nu \hat{n}_{\nu\uparrow} \hat{n}_{\nu\downarrow} + U_{12} \hat{n}_1 \hat{n}_2 + t_{12} \sum_s \left[ \hat{a}_{1s}^\dagger \hat{a}_{2s} + \text{H.c.} \right]. \tag{24}$$

Here, we define the occupancy operator $\hat{n}_{\nu s} = \hat{a}_{\nu s}^\dagger \hat{a}_{\nu s}$ for the spin-$s$ electron with the on-site energy $\epsilon_{\nu s}$ and $\hat{n}_\nu = \sum_s \hat{n}_{\nu s}$ for impurity-$\nu$ ($\nu = 1,2$). $U_\nu$ is the on-site Coulomb repulsion energy for impurity-$\nu$, and $U_{12}$ and $t_{12}$ are the inter-impurity Coulomb repulsion and hopping energies, respectively.

With the aid of the creation and annihilation operators, the spin operators associated with impurity-$\nu$ are expressed as $\widehat{\mathbf{S}}_\nu = 1/2 \sum_{s,s'} \hat{a}_{\nu s}^\dagger \boldsymbol{\sigma}_{ss'} \hat{a}_{\nu s'} = \{\hat{S}_{\nu x}, \hat{S}_{\nu y}, \hat{S}_{\nu z}\}$, and $\widehat{\mathbf{S}} = \sum_\nu \widehat{\mathbf{S}}_\nu = \{\hat{S}_x, \hat{S}_y, \hat{S}_z\}$ is the total spin operator of the multi-impurity system. Here, $\boldsymbol{\sigma} \equiv \{\sigma_x, \sigma_y, \sigma_z\}$ represents the vector of Pauli matrices. The spin-spin interaction Hamiltonian term

$$\widehat{H}^{\text{spin}} = J_{12} \widehat{\mathbf{S}}_1 \cdot \widehat{\mathbf{S}}_2 + \widetilde{D}_{12} (3\hat{S}_{1z}\hat{S}_{2z} - \widehat{\mathbf{S}}_1 \cdot \widehat{\mathbf{S}}_2) + \widehat{H}_S^{\text{soc}} \tag{25}$$

takes into account both the Heisenberg spin-exchange $J_{12}$ and the dipolar coupling $\widetilde{D}_{12}$ between two impurities. The last term describes the zero-field splitting for magnetic impurity systems with magnetic anisotropy induced by spin-orbit coupling (SOC) represented by

$$\widehat{H}_S^{\text{soc}} = D\hat{S}_z^2 + E(\hat{S}_x^2 - \hat{S}_y^2) \tag{26}$$

Here, $D$ and $E$ are the axial and in-plane MAE of the nanomagnet and can be evaluated via

$$D = D_{zz} - \frac{1}{2}(D_{xx} + D_{yy}), \quad E = \frac{1}{2}(D_{xx} - d_{yy}). \tag{27}$$

$\{D_{ii}\}$ with $i = x, y, z$ are the diagonal elements of the magnetic anisotropy tensor $\boldsymbol{D}$ which can be obtained from *ab initio* calculations.

In experiments, some external fields (such as the time-dependent gate voltage and magnetic field, etc) can be applied to local impurity system which leads to an additional external field term $\widehat{H}^{\text{ext}}$ in the system Hamiltonian. Consider a static magnetic field $\boldsymbol{B}_v$ and a time-dependent gate voltage $\tilde{\epsilon}_{vs}(t)$ both applied on impurity-$v$. The influence of such external fields reads

$$\widehat{H}^{\text{ext}}(t) = \sum_{v,s} \tilde{\epsilon}_{vs}(t)\hat{n}_{vs} + g\mu_B \boldsymbol{B}_v \cdot \widehat{\boldsymbol{S}}_v. \tag{28}$$

Here, $\tilde{\epsilon}_{vs}(t)$ modulates the impurity level $\epsilon_{vs}$ and results in charge fluctuations, while the magnetic field $\boldsymbol{B}_v \equiv \{B_v^x, B_v^y, B_v^z\}$ induces the Zeeman splitting and spin polarization. $g$ and $\mu_B$ denote the gyromagnetic factor and the Bohr magneton, respectively.

In HEOM-QUICK2, the fermionic environment is modelled by a collection of non-interacting electronic reservoirs whose correlation function includes all information of the environment. As reviewed in Sec. 2.3, the program has integrated high-performance spectrum decomposition schemes, which are applicable to environments covering a wide temperature range (e.g. from room temperature to liquid nitrogen and even liquid helium temperature). HEOM-QUICK2 has several preset types of time-dependent bias voltages i.e., the exponential voltage $V(t) = V_0[1 - \exp(-t/\tau_c)]$, the sinusoidal voltage $V(t) = V_0 \sin(2\pi t/\tau_c)$ and the delta-pulse voltage $V(t) = V_0 \delta(t)$. Here, $V_0$ measures the magnitude of the bias voltage and $\tau_c$ denotes the characteristic time, and $\delta(t)$ describes the duration of the pulse voltage (pulse width).

Carefully choosing proper system models and setting environmental parameters, HEOM-QUICK2 is capable of simulating realistic many-body OQSs involving diverse types of electron-electron and spin-spin interactions in the presence of external field. As an example, we employ the two-impurity Anderson model in Figure 9 to model the single molecular junction in the SP-STM setup. A static magnetic field along the $z$-direction is applied to each impurity. We explicitly take into account spin-specific impurity-reservoir coupling to characterize the influence of the spin-polarized tip on different spins of local impurities. The optimizer is applied to solve the stationary state of the impurity system and the hierarchy with the adiabatic terminator is truncated at $L = 3$.

### 4.2 Representative Applications of HEOM-QUICK2

The program evaluates a variety of important system properties in the stationary-state calculations:

- Spin-polarized electric current flowing into reservoir-$\alpha$

$$I_{\alpha s} \equiv -2 \sum_{\nu,p} \text{Im}\{\text{tr}_S[\hat{a}_{\nu s} \rho^+_{\alpha\nu p s}]\},$$

- Occupancy of spin-$s$ electrons on impurity-$\nu$

$$\langle \hat{n}_{\nu s} \rangle \equiv \text{tr}_S[\hat{n}_{\nu s}\, \rho^{(0)}],$$

- Spin moment of impurity-$\nu$ in different directions

$$\langle \hat{S}_{\nu,(x,y,z)} \rangle \equiv \text{tr}_S[\hat{S}_{\nu,(x,y,z)}\, \rho^{(0)}],$$

- Spin-spin correlation

$$\langle \hat{\mathbf{S}}_1 \cdot \hat{\mathbf{S}}_2 \rangle \equiv \text{tr}_S[\hat{\mathbf{S}}_1 \cdot \hat{\mathbf{S}}_2\, \rho^{(0)}],$$

- Internal energy of impurity

$$E_S \equiv \langle \hat{H}_S \rangle = \text{tr}_S[\hat{H}_S\, \rho^{(0)}],$$

- Impurity-reservoir hybridization energy

$$E_{\text{hyb}} \equiv \langle \hat{H}_{\text{hyb}} \rangle = 2 \sum_{\alpha,\nu,p,s} \text{Re}\{\text{tr}_S[\hat{a}_{\nu s} \rho^+_{\alpha\nu p s}]\}.$$

Some of them are exhibited in the output file of Figure 9. The program also supports the evaluation of the user-defined system properties.

HEOM-QUICK2 also allows for calculating the response properties of many-body OQSs, including

- Correlation function

$$C_{\hat{a}^\dagger_{\nu s} \hat{a}_{\nu s}}(t) \equiv \langle \hat{a}^\dagger_{\nu s}(t) \hat{a}_{\nu s}(0) \rangle,$$

- Lesser Green function

$$G^<_{\nu s}(\omega) = 2i\text{Re}\left\{\int_0^\infty C_{\hat{a}^\dagger_{\nu s} \hat{a}_{\nu s}}(t) e^{i\omega t} dt\right\},$$

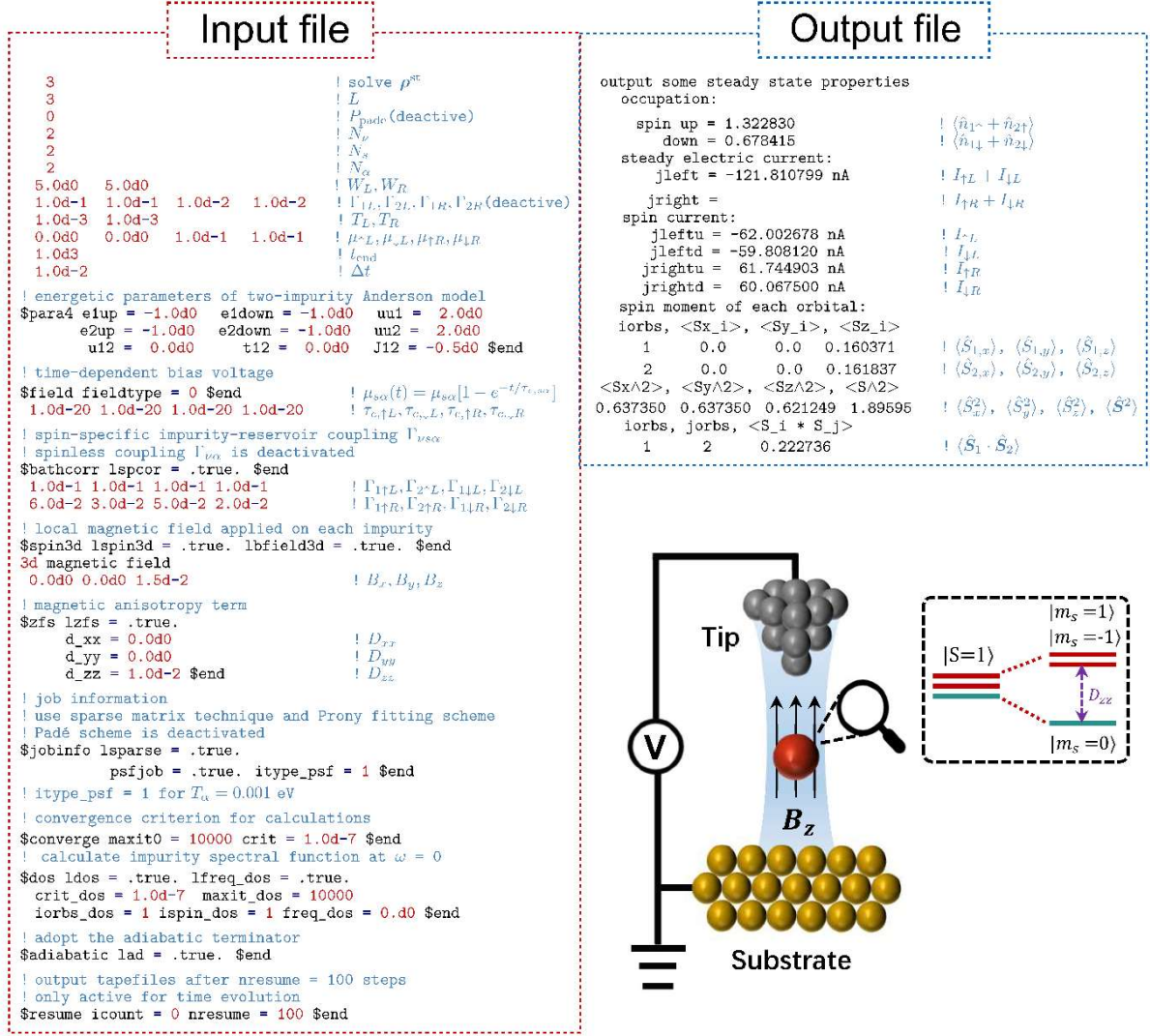

Figure 9: The left panel shows the standard input file for calculating the stationary state of a nanomagnet adsorbed on the substrate in the SP-STM setup. The molecular junction is described by the extended two-impurity Anderson model with local magnetic anisotropy. The influence of SP-tip on the local impurities is explicitly taken into account by the spin-specific system-reservoir couplings. A static magnetic field along the $z$-direction is applied to each impurity. The right panel exhibits some important system properties, including occupation number $\langle \hat{n}_{vs} \rangle$, spin-polarized current $I_{\alpha s}$, spin moment $\langle \hat{S}_{v,(x,y,z)} \rangle$ and spin correlation $\langle \hat{S}_1 \cdot \hat{S}_2 \rangle$, etc. The lower right panel displays the schematic of the SP-STM setup and an energy diagram for the local spin-triplet states of the nanomagnet in the absence of a magnetic field. The hierarchy with the adiabatic terminator is truncated at $L = 3$. The nonzero energetic parameters adopted are (in units of eV): $\epsilon_{vs} = -1.0$ for all $\{vs\}$, $U_1 = U_2 = 2.0$, $J_{12} = -0.5$, $D = 0.01$, $g\mu_B B_z = 0.015$, $\Gamma_{1\uparrow R} = 0.06$, $\Gamma_{2\uparrow R} = 0.03$, $\Gamma_{1\downarrow R} = 0.05$ and $\Gamma_{2\downarrow R} = 0.02$, $\Gamma_{vsL} = 0.01$ for all $\{vs\}$, $\mu_{sL} = 0$ and $\mu_{sR} = 0.1$ for all $\{s\}$, $T_\alpha = 0.001$ and $W_\alpha = 5.0$ for all $\{\alpha\}$. The program provides both the natural and the atomic units. In the natural unit, the electric current is in nA when the energy is in eV.

- Retarded Green function

$$G^r_{vs}(\omega) = -i \int_0^\infty \left\{ C_{\hat{a}^\dagger_{vs}\hat{a}_{vs}}(t) + \left[C_{\hat{a}_{vs}\hat{a}^\dagger_{vs}}(t)\right]^* \right\} e^{i\omega t} dt,$$

- Self-energy due to electron-electron interactions

$$\Sigma_{vs}(\omega) = \omega - \Sigma_B(\omega) - [G^r_{vs}(\omega)]^{-1},$$

with $\Sigma_B(\omega)$ being the reservoir-induced self-energy.

- Time-dependent spectral function

$$A_{vs}(\omega, t) = \frac{1}{2\pi} \int_{-\infty}^{+\infty} \langle \{\hat{a}_{vs}(t+\tau), \hat{a}^\dagger_{vs}(t)\} \rangle e^{i\omega\tau} d\tau.$$

Based on these observable as well as the response properties, one can characterize the stationary state or simulate the intricate dissipative dynamics of a variety of many-body OQSs, as demonstrated by the following applications.

*4.2.1 Application 1 -- Reproducing spin excitation signatures in differential conductance spectra*

Recent experiments have explored spin excitations for a variety of atomic or molecular junctions. Figure 10(a) illustrates the spin-flip excitations in a single molecular junction described by quantum impurity model. A spin-down electron flows from the substrate into an impurity (orbital), and meanwhile a spin-up electron on the same impurity (orbital) tunnels into the STM tip. This leads to spin-flip on the local impurity and thus triggers the excitation from the Zeeman state $|\uparrow\uparrow\rangle$ to $|\uparrow\downarrow\rangle$. To activate such inelastic spin excitations, external energy sources, such as a bias voltage, are generally required to overcome the energy gap between the two local Zeeman states. Consequently, the spin-flip excitations generally give rise to inelastic electron tunnelling signatures in the measured $dI/dV$ spectra, as manifested by the step-like lineshapes at finite bias voltages corresponding to the excitation energy.

The experimentally measured spin excitation energies $\Delta E$ can, indeed, be as low as a few meV or even at the two sub-meV level[117,221-223]. This energy scale is several orders of magnitude lower than those associated with charge transfer between the system and its environment. HEOM-QUICK (version 1) combined with the *ab initio* approach has been employed to study the inelastic tunnelling signatures of such low-energy spin excitations in realistic single molecular junctions[140,161,162,190]. However, the values of $\Delta E$ have to be scaled up by several times or even orders of magnitude, because a sufficiently low temperature must be adopted in the simulation to resolve the step-like lineshape of in the $dI/dV$ spectra.

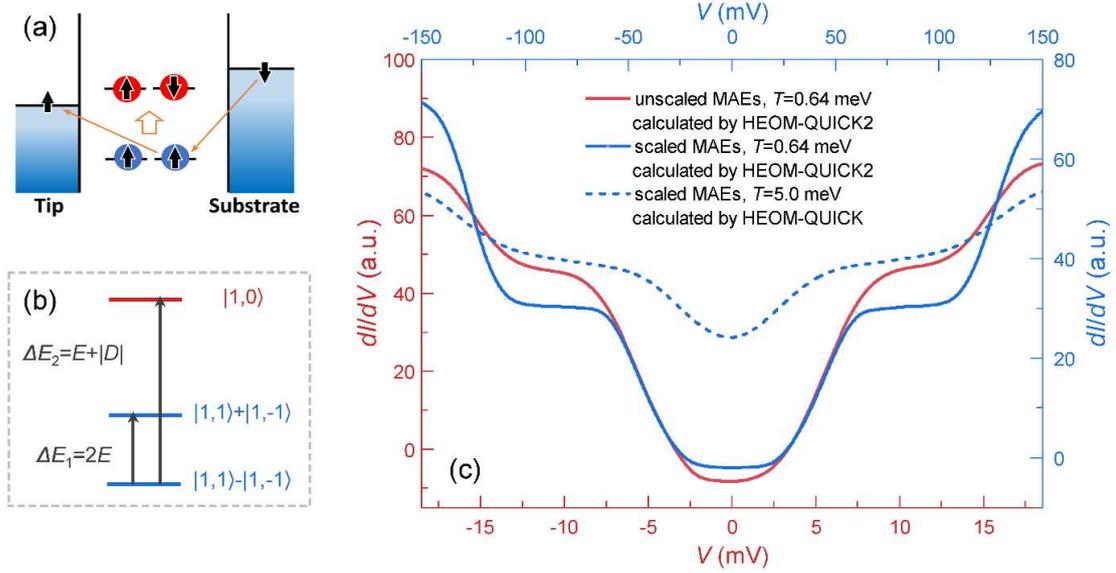

Figure 10: The revisit of the differential conductance spectra of a dehydrogenated FePc/Au(111) junction[192] (Copyright from American Institute of Physics in 2023 with permission). Panel (a) illustrates the inelastic electron cotunneling process responsible for the spin-flip excitation from the Zeeman state $|\uparrow\uparrow\rangle$ to $|\uparrow\downarrow\rangle$. Panel (b) gives an energy diagram for the local spin-triplet states on the molecular magnet. Here, the local spin states of the system are denoted as $|S, m_s\rangle$. The $dI/dV$ spectra in (c) are calculated by HEOM-QUICK (version 1) (dash lines) and HEOM-QUICK2 (solid lines) for a two-impurity Anderson model with the scaled MAE ($D = -102.7$ meV and $E = 2.2$ meV) and unscaled MAE ($D = -13.0$ meV and $E = 2.2$ meV), respectively. The blue lines use the upper and right axes, and the red lines use the bottom and left axes. The numerical data [blue dashed line in (c)] obtained by HEOM-QUICK (version 1) are extracted from Ref. [162] (Copyright 2019 American Institute of Physics). The hierarchy is truncated at $L = 3$ in HEOM-QUICK2 calculations. The other energetic parameters are (in units of eV): $\epsilon_{1s} = \epsilon_{2s} = -2.5$ for all $\{s\}$, $U_1 = U_2 = 5$, $\Gamma_{1s} = 0.06$ and $\Gamma_{2s} = 0.1$ for all $\{s\}$, $\Gamma_{vsL} = 0.01$ for all $\{vs\}$, $T_\alpha = 0.00064$ and $W_\alpha = 5.0$ for all $\{\alpha\}$.

To demonstrate the improvement of HEOM-QUICK2, we revisit the previous simulation by HEOM-QUICK (version 1) for the $dI/dV$ spectra of a dehydrogenated iron phthalocyanine (FePc) molecules adsorbed on Au(111) surface[162]. The FePc/Au(111) junction is described by the extended two-impurity Anderson model with two electron reservoirs. The axial and in-plane MAE of the FePc molecule take the unscaled values measured in the STM experiment, i.e., $D = -13.0$ meV and $E = $ 2.2 meV[224]. Figure 10(b) depicts the energy diagram and corresponding excitation energies among the local spin-triplet states of the molecular magnet.

Figure 10(c) depicts the $dI/dV$ spectra calculated for the scaled and unscaled MAE, respectively. With the unscaled MAE, an overall symmetric lineshape with four steps is clearly exhibited at $V_{1,\pm} = \pm 5$ mV and $V_{2,\pm} = \pm 15$ mV, respectively. Each step corresponds to a voltage-driven spin excitation within the single molecular junction and the voltage amplitudes $|V_{1,\pm}|$ and $|V_{2,\pm}|$ agree closely with the two theoretical excitation energies, $E_1 = 2E$ and $E_2 = |D| + E$, respectively. The $dI/dV$ spectra calculated with the scaled MAE are also shown in Figure 10(c). The overall lineshape and positions of the steps are similar to the spectrum calculated with the unscaled MAE, given the same scaling is applied to the voltage.

*4.2.2 Application 2 -- Simulating the long-time formation of Kondo states*

Exploring real-time dynamics of strongly correlated states in many-body OQSs holds fundamental importance for understanding the mechanisms behind numerous exotic quantum phenomena and for realizing on-demand quantum control. Recent experimental advancement has enabled the precise measurement on the evolution of local electronic states, particularly under the influence of time-dependent external fields and dissipative environment[176,225-232].

To demonstrate the numerical performance of HEOM-QUICK2 for simulating the dissipative dynamics, we explore the spin relaxation process of a strongly correlated OQS described by the single Anderson impurity model, which was previously studied by the time-dependent NRG (TDNRG) approach[15]. The two-step dynamic simulation proceeds as follows:

1) Initially, a free impurity is decoupled from the reservoir, i.e., $\Gamma = 0$. The impurity levels of spin states $|\uparrow\rangle$ and $|\downarrow\rangle$ are populated at the Fermi level $\epsilon_\uparrow = 0$ and $\epsilon_\downarrow = \Gamma_0$, respectively. Carry out the stationary-state calculation to obtain the initial state for the subsequent time evolution.

2) At time $t = t_0$, the system-reservoir coupling is switched on, i.e., $\Gamma = \Gamma_0$, and the impurity energy levels are simultaneously shifted to $\epsilon_\uparrow = \epsilon_\downarrow = -U/2$. Employ the dynamic simulator to propagate the systems freely from the prepared initial state until the final time $t_f$. The program evaluates and records the spin properties of the system.

For a wide range of impurity-reservoir coupling strength, the resulting time evolution of magnetization $\langle \hat{S}_z \rangle = 1/2 \langle \hat{n}_\uparrow - \hat{n}_\downarrow \rangle$ in Figure 11(a) and (b) agrees closely with the original TDNRG results shown in Figure 11(c) and (d). Such an agreement highlights the numerical accuracy and reliability of HEOM-QUICK2 to investigate the long-time dissipative dynamics of strongly correlated OQSs. Moreover, two distinct time scales are clearly visible for the spin relaxation. Scaling time with $1/\Gamma$ yields a universal short-time relaxation, and the long-time relaxation is governed by the Kondo time scale $1/T_K$. Here, the Kondo temperature $T_K$ is evaluated by $T_K = \sqrt{\Gamma U/2}\, e^{-\pi U/8\Gamma + \pi\Gamma/2U}$ for the symmetric Anderson model[170].

We now consider the continuous weakening of the Kondo state triggered by a switch-on spin-exchange interaction in quantum impurity systems coupled to a single reservoir. The system is described by a two-impurity Anderson model of Equation (24). To monitor the evolution of the Kondo state, we examine the full spectrum of time-dependent impurity spectral functions $A_{vs}(\omega, t)$ at several carefully selected time instants. The spectral function provides a comprehensive description of the density of states of the impurity system. The height of Kondo peak residing in the low energy region characterizes the strength of Kondo correlation. The calculation proceeds in three steps as follows:

1) Initially, there is no inter-impurity spin-exchange interaction in the impurity system. Solve the stationary state of the system and take it as initial state for the subsequent dynamic simulation.

2) At time $t = t_0$, the ferromagnetic spin-exchange interaction $\hat{V} = J_{12}\, \hat{\mathbf{S}}_1 \cdot \hat{\mathbf{S}}_2$ switches on. Propagate the impurity system freely from the initial state until the final time $t_f$. The program records the intermediate $\boldsymbol{\rho}(t)$ at chosen time instants.

3) Employ $\boldsymbol{\rho}(t)$ to evaluate the time-dependent impurity spectral function $A_{vs}(\omega, t)$ at each chosen time instant.

Figure 12 exhibits the relaxation dynamics of Kondo states. Before imposing the spin-exchange interaction $\hat{V}$ at time $t = t_0$, the local spin states of each impurity are fully screened by the conducting electrons in the reservoir. It is evidenced by the sharp resonance peak around $\omega = 0$ in the full spectra of $A_{1\uparrow}(\omega, t)$; see Figure 12(b). Driven by the ferromagnetic interaction $\hat{V}$, the initially separated local impurity spins undergo a merging process, gradually forming a local high-spin state during the relaxation process. Subsequently, the electron spins in the reservoir realign themselves to screen the local high spin. However, such screening is not complete due of the insufficient screening channels, and thus the dissipative relaxation ultimately yields a underscreened Kondo states[233,234]. Consequently, the Kondo correlation associated with each impurity is weakened continually, as indicated by the progressively reduced resonance peak centered at $\omega = 0$ of the spectral function $A_{1\uparrow}(\omega, t)$ at the chosen time instants in Figure 12(b).

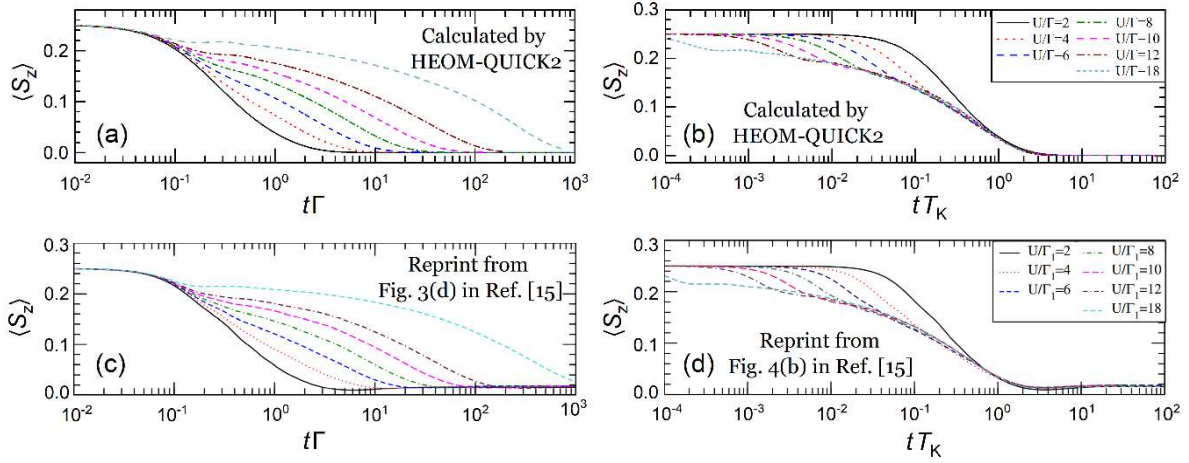

Figure 11: Time-dependent magnetization $\langle \hat{S}_z \rangle$ of the single impurity Anderson model with different Coulomb interaction energy $U$. The panels (a) and (b) depict results calculated by HEOM with the adiabatic terminator truncated at $L = 4$, while the panels (c) and (d) display the time-dependent NRG (TDNRG) results extracted from Ref. [15] (Copyright from American Physical Society in 2005 with permission). In the left panels the magnetization curves are plotted verse $t\Gamma$ and in the right panels the curves are replotted verse $tT_K$. The Kondo temperature is evaluated by $T_K = \sqrt{\Gamma U/2}\ e^{-\pi U/8\Gamma + \pi\Gamma/2U}$[170]. The results of HEOM-QUICK2 agree closely with the zero-temperature results obtained by the TDNRG method.

*4.2.3 Application 3 -- Simulating the long-time formation of spin-spin correlation*

We employ HEOM-QUICK2 to simulate the formation of the spin states of a single molecular junction in a STM setup[192]. The junction is described by the two-impurity Anderson model with a non-zero inter-impurity Heisenberg exchange interaction which gives rise to an energy gap $\Delta E = |J|$ between the singlet and triplet states of the molecular magnet. The time evolution of local spin states is characterized by the square of spin angular momentum $\langle \hat{S}^2 \rangle = \text{tr}_S[\hat{S}^2 \rho_S]$ and the time-dependent impurity spectral functions $A_{\nu s}(\omega, t)$, respectively. The calculation is performed in three steps as follows:

1)  Solve the stationary state of the system in the absence of a bias voltage and take it as the initial state for the subsequent time evolution;

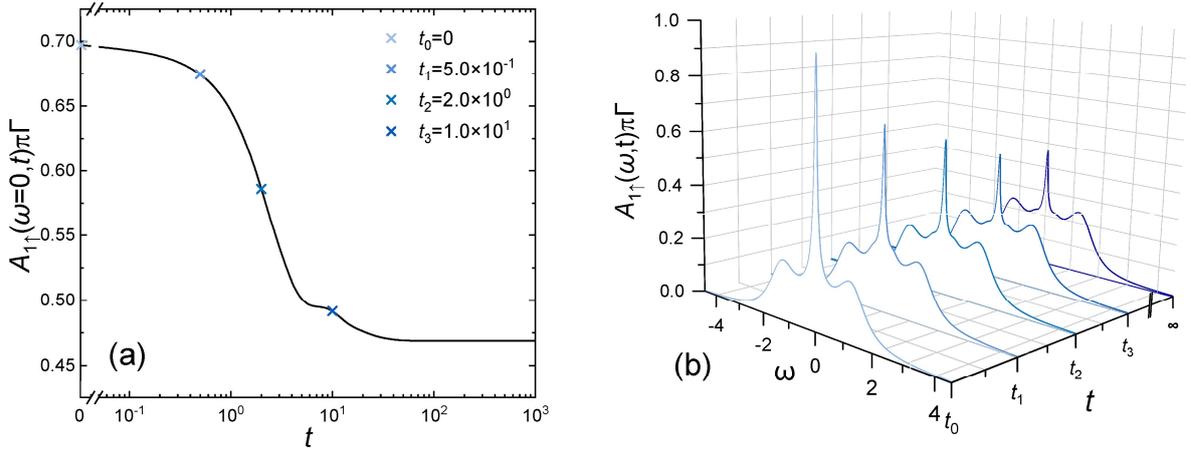

Figure 12: (a) Time-dependent impurity spectral functions $A_{1\uparrow}(\omega, t)$ calculated for a two-impurity Anderson model with ferromagnetic exchange interaction. Specifically, $t = t_0$ is the time instants when ferromagnetic interaction switches on immediately between the two impurities. Panel (b) shows the full spectrum of $A_{1\uparrow}(\omega, t)$ at a number of time instants (marked by coloured crosses in the panel (a)). The stationary-state impurity spectral function is taken as the result of $A_{1\uparrow}(\omega, t)$ at $t \to \infty$. The hierarchy with the adiabatic terminator is truncated at $L = 4$. The nonzero energetic parameters adopted are (in units of eV): $\epsilon_{vs} = -1.0$ for all $\{vs\}$, $U_1 = U_2 = 2.0$, $J_{12} = -0.5$, $\Gamma_{vs} = 0.6$ for all $\{vs\}$, $T_\alpha = 0.001$ and $W_\alpha = 4.0$ for all $\{\alpha\}$.

2) Impose a bias voltage $|V| > \Delta E$ across the reservoirs at time $t_0$ to activate the spin excitations between the singlet and triplet states. The program records the intermediate $\rho(t)$ at chosen time instants;

3) Employ $\rho(t)$ to evaluate the time-dependent impurity spectral function $A_{vs}(\omega, t)$ at each chosen time instant.

Figure 13(a) depicts the time evolution of the local spin states. Before switching on the bias voltage at $t_0$, the system stays in a spin-triplet state with $\langle \hat{S}^2 \rangle = 1.97$. This is justified by Figure 13(b), where a single peak at $\omega \approx \epsilon_0$ emerges in the spectral function $A_{1\uparrow}(\omega, t)$, corresponding to the local spin-triplet state. During the subsequent dissipative dynamics of the system, the spin excitation from the triplet state to the singlet state is activated by the bias voltage, as indicated by the decrease of $\langle \hat{S}^2 \rangle$ to the final value of 1.76. Meanwhile the intensity of the main spectral peak is weakened gradually; while the satellite peak corresponding to the spin excited state emerges at $\omega \approx \epsilon_0 + |J|$ and becomes prominent; see Figure 13(b).

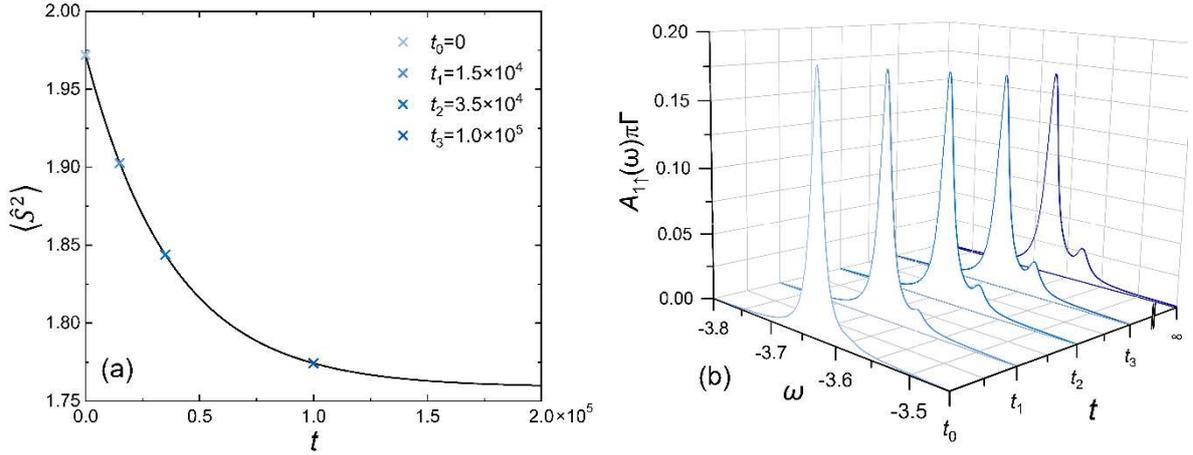

Figure 13: (a) Time evolution of $\langle \hat{S}^2 \rangle$ for the two-impurity Anderson model driven by a bias voltage switched on at $t = t_0$[192]. Panel (b) shows $A_{1\uparrow}(\omega, t)$ at different time instants (marked by coloured crosses in the panel (a)). The stationary-state impurity spectral function is taken as the result of $A_{1\uparrow}(\omega, t)$ at $t \to \infty$. The hierarchy is truncated at $L = 3$. The nonzero energetic parameters adopted are (in units of eV): $\epsilon_{\nu s} = -3.6$ for all $\{\nu s\}$, $U_1 = 4.6$, $U_2 = 4.5$, $J_{12} = -0.05$, $\Gamma_{\nu sL} = 0.005$ and $\Gamma_{\nu sR} = 0.0025$ for all $\{\nu s\}$, $\mu_{sL} = 0$ and $\mu_{sR} = -0.2$ for all $\{s\}$, $T_\alpha = 0.001$ and $W_\alpha = 5.0$ for all $\{\alpha\}$ (Copyright from American Institute of Physics in 2022 with permission).

**Figures, Video, Audio and Tables**

**Sidebar title:**

[Please include sidebars in the body of the text where appropriate]

**Conclusion**

We have reviewed the recent theoretical and numerical efforts on the development of fermionic HEOM method, which gives rise to a general-purpose simulator HEOM-QUICK2 for fermionic many-body OQSs. The code architecture has been introduced in detail. This is followed by the representative applications that exemplify the high performance of HEOM-QUICK2. This program features more efficient solvers for stationary states, more accurate treatment of the non-Markovian environmental memory, and the improved numerical convergence and stability for longtime dissipative dynamics. Furthermore, the unprecedented precision achieved by HEOM-QUICK2 enables accurate simulations of low-energy spin excitations and coherent spin relaxation processes. The development of HEOM-QUICK2 represents a significant advancement in the field of quantum statistical dynamics, offering a powerful and comprehensive tool for studying many-body OQSs. We believe that it is promising for HEOM-QUICK2 to set precise benchmarks for other quantum dynamics software for studying fermionic many-body OQSs.

The primary challenge that HEOM encounters is the "exponential wall" in the treatment of complex many-body OQSs with a large $N$. To overcome this challenge and make substantial advancements in future developments, it is crucial to combine the HEOM method with cutting-edge numerical algorithms. This integration is essential for effectively reducing the computational costs, thus paving the way to "break the wall" and ultimately achieving high-performance computing for complex many-body OQSs. Recently, the low-rank tensor decomposition has aroused a continuously growing interest. One representative example is the matrix product state (MPS) technique, which decomposes the higher-order tensors with multiple indices into a product form of the lower-order tensor with fewer indices. Several studies have successfully combined the HEOM method with the MPS technique, enabling simulations of dissipative dynamics in many-body OQSs[88,91,94,235-239]. The developments of artificial intelligence[240-244] and quantum computing techniques[245-248] have also provided new insights into the dynamics of many-body OQSs. It remains to explore novel HEOM theoretical frameworks applicable to artificial neural networks and quantum computing.

Recently, the ESR spectroscopy for single molecules has been realized in the STM setup by invoking the radio-frequency voltages. The STM-ESR technique has greatly improved the energy resolution (<1 $\mu$eV) in the measurement of spin-spin interactions[176,177]. The integration of STM with pump-probe techniques has enabled the time-resolved measurement and control of local spin states[229,231,232]. This experimental progress also presents new challenges to the numerical precision of the HEOM method.

In future, our attention will be devoted to the combination of the HEOM method and the MPS technique to explore more intricate dissipative dynamics in many-body OQSs. We are also attempting to construct fermionic HEOM by using the artificial neural network techniques and quantum computing algorithms to accelerate the HEOM computation. Furthermore, we plan to carry out a systematic benchmark on the performance among latest efficient truncation schemes and iterative algorithms[199-201,218-220,249-251] and plant the superior ones in a future version of our program. These approaches are expected to substantially reduce computational cost and extend the applicability of HEOM to the future frontier of theoretical and experimental research, including the time-resolved measurement and manipulation of spin states, simulating the signal of STM-ESR spectroscopy with extremely high energy resolution and the on-demand control of spin qubits at the atomic scale.

**Funding Information**

The support from the National Natural Science Foundation of China (Grant Nos. 22321003, 22393912, 22103073, 22203083 and 22173088), the Strategic Priority Research Program of the Chinese Academy of Sciences (Grant No. XDB0450100), and the Innovation Program for Quantum Science and Technology (Grant No. 2021ZD0303301) is gratefully acknowledged.

**Acknowledgments**

The computational resources are provided by the Supercomputing Center of University of Science and Technology of China. The authors are indebted to Hou-Dao Zhang and Zi-Hao Chen for their invaluable help.

**Notes**

The HEOM-QUICK2 program is openly available on GitHub at: https://github.com/zdcwork/HEOM-QUICK2, and the open source data that supports the findings of this study can be found on GitHub at: https://github.com/zdcwork/open-source-data.